\documentclass[pdflatex,sn-mathphys-num,iicol]{sn-jnl}

\usepackage{hyperref}    
\usepackage[italic]{belle2-particles}
\usepackage{belle2-symbols}
\usepackage{amsfonts}
\usepackage{amsmath} 
\usepackage{amssymb}
\usepackage{array}
\usepackage{blindtext} 
\usepackage{bm}
\usepackage{caption}
\usepackage[noabbrev,capitalise]{cleveref}
\usepackage{color}
\usepackage{colortbl}
\usepackage{floatrow}
\usepackage{hypcap} 
\usepackage{ifpdf}
\usepackage{graphicx}  
\usepackage{highlight} 
\usepackage{longtable}
\usepackage{lineno}  
\usepackage{mathrsfs}
\usepackage{mciteplus}
\usepackage[version=4]{mhchem}
\usepackage{microtype}
\usepackage{multirow}
\usepackage{overpic}
\usepackage{pgffor} 
\usepackage{physics}
\usepackage{relsize}
\usepackage{rotating}
\usepackage[separate-uncertainty]{siunitx} 
\usepackage{subfigure}
\usepackage{tikz}
\usepackage{xfp} 
\usepackage{xspace} 
\usepackage{upgreek} 
\usepackage{url}

\usepackage{forloop}
\usepackage{orcidlink}

\graphicspath{{./figs/}} 

\newcommand*\patchAmsMathEnvironmentForLineno[1]{%
\expandafter\let\csname old#1\expandafter\endcsname\csname #1\endcsname
\expandafter\let\csname oldend#1\expandafter\endcsname\csname
end#1\endcsname
 \renewenvironment{#1}%
   {\linenomath\csname old#1\endcsname}%
   {\csname oldend#1\endcsname\endlinenomath}%
}
\newcommand*\patchBothAmsMathEnvironmentsForLineno[1]{%
  \patchAmsMathEnvironmentForLineno{#1}%
  \patchAmsMathEnvironmentForLineno{#1*}%
}
\AtBeginDocument{%
\patchBothAmsMathEnvironmentsForLineno{equation}%
\patchBothAmsMathEnvironmentsForLineno{align}%
\patchBothAmsMathEnvironmentsForLineno{flalign}%
\patchBothAmsMathEnvironmentsForLineno{alignat}%
\patchBothAmsMathEnvironmentsForLineno{gather}%
\patchBothAmsMathEnvironmentsForLineno{multline}%
}

\newfloatcommand{capbtabbox}{table}[][\FBwidth]

\ExplSyntaxOn
\RenewDocumentCommand \SIrange { O{} m m m } {
  \leavevmode
  \group_begin:
     \keys_set:nn { siunitx } { open-bracket = [,  close-bracket= ], #1}
     \__siunitx_range_unit:nnnn {#4} { open-bracket = [,  close-bracket= ], #1} {#2} {#3}
  \group_end:
}
\ExplSyntaxOff

\newcommand{\twocolumnplotwidth}{0.49\textwidth}

\newcommand{\Sup}[1]{\ensuremath{^{\text{#1}}}\xspace}

\let\sub\Sub 

\NewDocumentCommand{\dEdx}{s}{\ensuremath{\IfBooleanTF{#1}{\dv*{E}{x}}{\dv{E}{x}}}\xspace}

\theoremstyle{thmstyleone}%
\theoremstyle{thmstyletwo}%

\theoremstyle{thmstylethree}%

\begin{document}

\title[Article Title]{Charged-hadron identification at \BelleII}

\author{
   I.~Adachi\,\orcidlink{0000-0003-2287-0173},  
   H.~Ahmed\,\orcidlink{0000-0003-3976-7498},  
   Y.~Ahn\,\orcidlink{0000-0001-6820-0576},  
   H.~Aihara\,\orcidlink{0000-0002-1907-5964},  
   N.~Akopov\,\orcidlink{0000-0002-4425-2096},  
   A.~Albert\,\orcidlink{0000-0002-1251-0564},  
   S.~Alghamdi\,\orcidlink{0000-0001-7609-112X},  
   M.~Alhakami\,\orcidlink{0000-0002-2234-8628},  
   A.~Aloisio\,\orcidlink{0000-0002-3883-6693},  
   N.~Althubiti\,\orcidlink{0000-0003-1513-0409}
% ...
\\ 
   {\centering (The Belle II Collaboration)}
}

\abstract{The \BelleII experiment's ability to identify particles critically affects the sensitivity of its measurements.
We describe \BelleII's algorithms for identifying charged particles and evaluate their performance in separating pions, kaons, and protons using \SI{426}{fb^{-1}} of data collected at the energy-asymmetric $\APelectron\Pelectron$ collider SuperKEKB in 2019--2022 at center-of-mass energies at and near the mass of the \PUpsilonFourS.}




\maketitle

\section*{Introduction\addcontentsline{toc}{section}{Introduction}}
\label{sec:intro}

The \BelleII experiment is located at the energy-asymmetric $\APelectron\Pelectron$ collider SuperKEKB~\cite{Akai:2018mbz} in Tsukuba, Japan.
It began taking data in March 2019 and aims to accumulate \SI{50}{ab^{-1}} of integrated luminosity at center-of-mass~(c.m.) energies at and near the mass of the \PUpsilonFourS.
It has broad goals such as quantifying \CP violation in heavy-meson decays, measuring the parameters of the Cabibbo-Kobayashi-Maskawa quark-mixing matrix, studying exotic particles with spectroscopy, determining the properties of the \Ptau lepton, and searching for particles and forces beyond those of the standard model~\cite{Belle-II:2018jsg}. Its physics program is an expansion of the \mbox{\slshape B\kern-0.1em{\smaller A}\kern-0.1em B\kern-0.1em{\smaller A\kern-0.2em R}} and Belle experiments~\cite{Bevan:2014iga}, which together accumulated \SI{1.5}{ab^{-1}} of data, but with new challenges stemming from its much higher instantaneous luminosity and background rate. 

At \BelleII, we detect six species of charged particles: electrons, muons, pions, kaons, protons, and deuterons.
It is important that we distinguish them from each other, especially in the momentum range of the decay products of \PB and \PD mesons and \Ptau leptons, [0.1, 5.0] GeV/c.
Particle identification~(PID) is especially important for \PB-flavor tagging~\cite{Belle-II:2021zvj,Belle-II:2024lwr}, charm-flavor tagging~\cite{Belle-II:2023vra}, and full reconstruction of \PB mesons~\cite{Keck:2018lcd}.

We introduce the detector and data sets in sections~\ref{sec:detector} and \ref{sec:data}, define the PID likelihoods for individual detector components in section~\ref{sec:pid_model}, and describe how they are combined to identify particles in section~\ref{sec:selectors}. In sections~\ref{sec:pid_efficiency_and_misID_rate} and \ref{sec:performance}, we describe the analysis method and software infrastructure and evaluate performance for identifying pions, kaons, and protons (we do not yet have suitable control samples for deuterons) using control channels in \SI{426}{fb^{-1}} of data collected in 2019--2022. We conclude in section~\ref{sec:future} and outline development plans for the next years.\section{Detector}
\label{sec:detector}

\BelleII is a general-purpose detector consisting of seven subdetectors and a superconducting solenoid arranged cylindrically around the $\APelectron\Pelectron$ interaction region~\cite{Belle-II:2018jsg, Abe:2010sj}.
From innermost to outermost, these subdetectors are the pixel vertex detector~(PXD), silicon vertex detector~(SVD), central drift chamber~(CDC), time-of-propagation detector~(TOP) and aerogel ring-imaging Cherenkov detector~(ARICH), electromagnetic calorimeter~(ECL), and \PKlong and muon detector~(KLM).
The solenoid, located between the ECL and the KLM, provides a \SI{1.5}{T} magnetic field nearly parallel to the beam directions.

The PXD consists of two layers of DEPFET pixel sensors, the first covering the full azimuthal range and the second only 20\% (the full second layer was installed in 2023).
The SVD~\cite{Adamczyk_2022} consists of four layers of double-sided silicon strip sensors.
The CDC, the main tracking subdectector, is a large volume of helium and ethane gas crossed by sense and field wires.
The TOP~\cite{Atmacan:2025jmh}, covering the barrel of the CDC, and the ARICH, covering its forward end cap, measure Cherenkov light produced by charged particles.
The TOP consists of quartz bars that internally reflect light, which is detected by micro-channel-plate photo-multiplier tubes.
The ARICH~\cite{Yonenaga:2020eby} consists of two layers of aerogel tiles, with different refractive indices, that focus light into sharp rings, detected by hybrid avalanche photon detectors.
The ECL covers the barrel and forward and backward end caps with thallium-doped cesium-iodide crystals, each \num{16.2} radiation lengths deep.
The KLM serves as the return yoke of the magnetic field, with gaps in the steel structure instrumented with scintillator strips in the end caps and first two layers of the barrel and with resistive plate chambers in the other barrel layers.

We define $\hat{z}$ as the cylindrical axis of the solenoid, with its positive direction nearly coincident with electron-beam direction (the beams collide with a crossing angle of \SI{83}{mrad}). Polar angles are defined relative to $\hat{z}$, and azimuths relative to the direction orthogonal to $\hat{z}$ that points outside the accelerator ring.
Our PID uses likelihoods that depend on a particle's momentum, $\vec{p}$, measured from the curvature of the particle's track in the magnetic field, which is reconstructed from the locations of hits in the PXD, SVD, and CDC;
$p$, $\theta$, and $\phi$ are the magnitude, polar angle, and azimuth of this momentum.
When a reference frame is not explicitly mentioned, variables are defined in the laboratory frame.

The origin of the coordinate system is the point at which the beams are expected to collide.
The $\APelectron\Pelectron$ interaction region depends on the data-taking period and is determined from $\APelectron\Pelectron\to\APmuon\Pmuon$ events.
It is usually within a millimeter of the origin.
\section{Data}
\label{sec:data}

We use data collected in 2019--2022 and simulation samples that resemble it to develop and study the PID likelihoods and their performance.
We simulate the detectors using their conditions as recorded during data taking and overlay background signals taken from randomly triggered events. 
We simulate the production of quark-antiquark and lepton-antilepton pairs from the $\APelectron\Pelectron$ collision with \textsc{KKMC}~\cite{Jadach:1999vf}, hadronization with \textsc{PYTHIA~8}~\cite{Sjostrand:2014zea}, hadron decay with \textsc{EvtGen}~\cite{Lange:2001uf}, and \Ptau decay with \textsc{Tauola}~\cite{Davidson:2010rw}.
For each charged-particle species, we also generate simulated data with particles isotropically distributed in the detector and evenly distributed in the range of momenta produced at \BelleII.
Detector response is simulated with \textsc{GEANT4}~\cite{GEANT4:2002zbu}, and we reconstruct both real and simulated data using the Belle II analysis software framework, \basfii~\cite{Kuhr:2018lps,basf2-zenodo}.


To develop, test, and evaluate our particle identification algorithms, we use data from control channels in which particles can be identified by context.
We reconstruct these channels without PID information, letting us calibrate PID from real data.
To evaluate hadron-ID performance, it is especially important to gather pure samples of pions, kaons, and protons,
both positively and negatively charged.
We use two-body \PDzero, \PKzero, and \PLambda decays to charged particles to isolate such samples.
In such decays, the negative \PDzero decay product is almost always a kaon and the positive one almost always a pion;
both \PKzero decay products are almost always pions;
and one \PLambda decay product is always a proton and the other almost always a pion.
To simplify descriptions throughout, we include charge-conjugated states without explicitly writing them.

We reconstruct $\PDzero \to \PKminus\Ppiplus$ and determine the \PDzero flavor by requiring the \PDzero be produced from \PDstarplus decay, $\PDstarplus \to \PDzero \Psoftpiplus$, where \Psoftpiplus is a soft pion, which has low momentum due to the small mass difference between a \PDstarplus and a \PDzero.
We pair \PDzero candidates with charged tracks, assumed pions, to form \PDstarplus candidates.
We assume the \PDzero decay product with the same charge as the soft pion is a pion and that with the opposite charge is a kaon.\footnote{The doubly-Cabibbo-suppressed decay $\PDzero \to \PKplus \Ppiminus$ has a branching fraction that is 0.4\% of the Cabibbo favored decay, $\PDzero \to \PKminus \Ppiplus$. Candidates reconstructed with the wrong particle assignment are generally correctly identified as background in the fit described below, reducing this contamination to below $0.1\%$.}
For each candidate decay chain, we require the reconstructed \PDzero mass be in the range [1.80, 1.95] GeV/c$^2$, the difference of the \PDstarplus and \PDzero masses be in the range [143.9, 146.9] MeV/c$^2$, and the \PDstarplus momentum in the c.m.\ frame be greater than~\SI{2.5}{GeV/c}.

For $\PKzero\to\Ppiplus\Ppiminus$, we assume both decay products are pions.
For each track, the $z$ coordinate of the point of closest approach to the origin must be less than \SI{4}{cm}.
We require that the mass of the reconstructed \PKzero be in the range [470, 530] MeV/c$^2$ and remove contamination from $\PLambda \to \Pproton \Ppiminus$ and photon conversion, $\Pphoton \to \APelectron\Pelectron$, by rejecting any candidate whose mass is in the range [1.11, 1.12] GeV/c$^2$ when we assume that either decay product is a proton or less than~\SI{50}{MeV/c^2} when we assume both decay products are electrons.
For each \PKzero candidate, we fit a common production vertex to information from both tracks and require the $\chi^2$ probability of the fit be above 1\%, the vertex be less than \SI{3.5}{cm} from the origin, and (to suppress random combinations of prompt tracks) the proper decay time be greater than~\SI{0.007}{ns}.

For each \PLambda candidate, we assume that the decay product with higher momentum is a proton and the other is a pion.
We require that the mass of the reconstructed \PLambda be in the range [1.105, 1.128] GeV/c$^2$ and the momentum of the proton be greater than~\SI{0.5}{GeV/c} and be~60--100\% of the \PLambda's momentum.
We remove contamination from \PKzero decay and photon conversion by rejecting any candidate whose mass is in the range [480, 520] MeV/c$^2$ when we assume that both decay products are pions or less than~\SI{50}{MeV/c^2} when we assume both decay products are electrons.
For each candidate, we fit both tracks to a common production vertex and require that its displacement vector from the interaction point be within~\SI{8}{\degree} of the \PLambda momentum vector.

\begin{figure*}[t!]
    \begin{center}

    \begin{tabular}{c c}
        \includegraphics[width=0.4\textwidth]{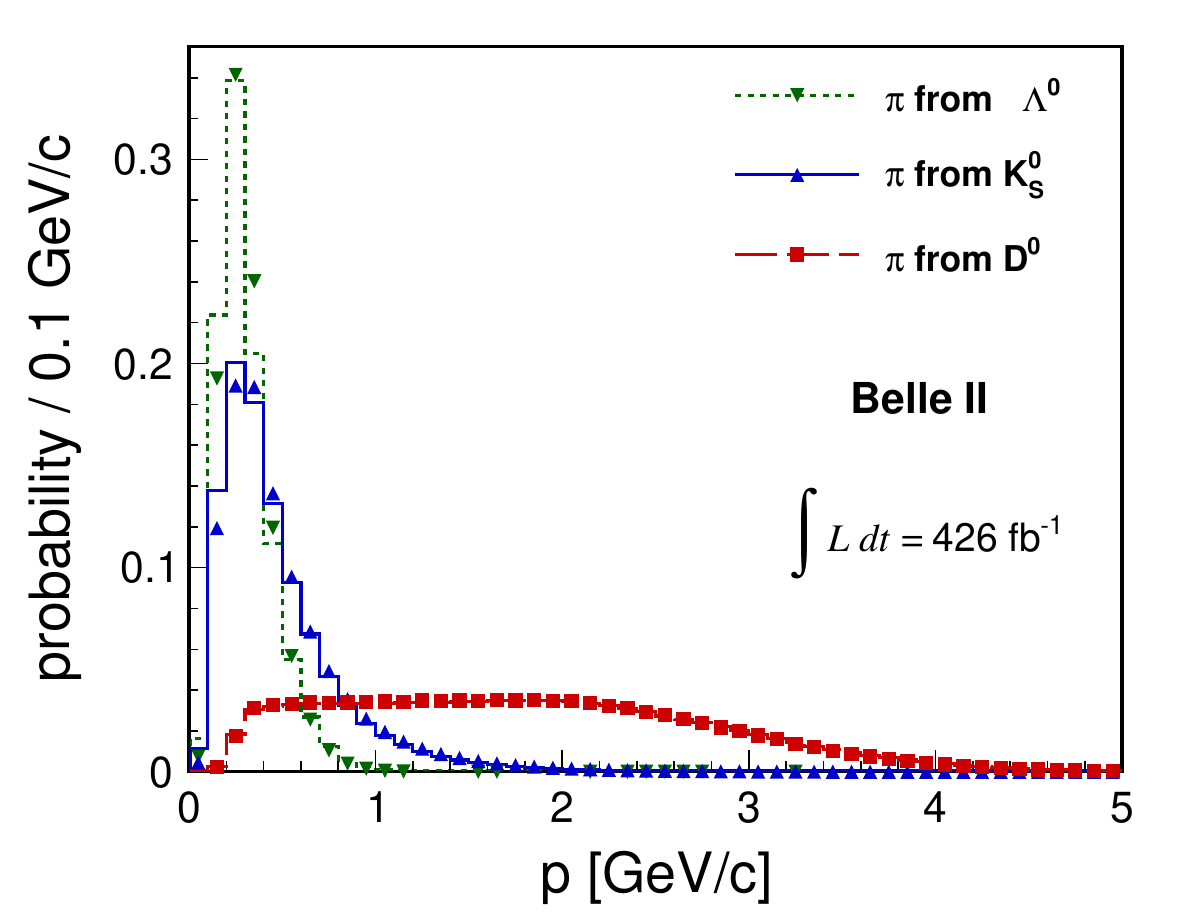} & 
        \includegraphics[width=0.4\textwidth]{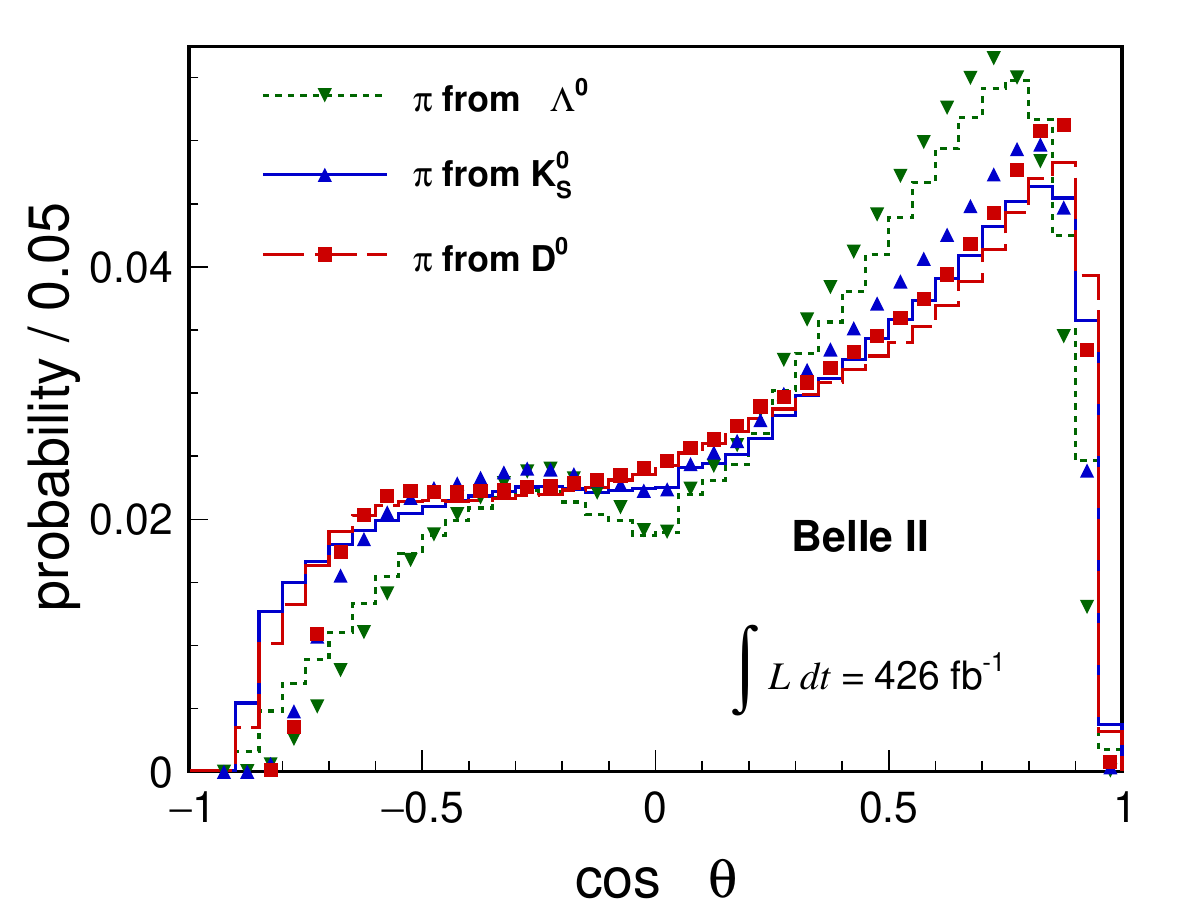} \\
        \includegraphics[width=0.4\textwidth]{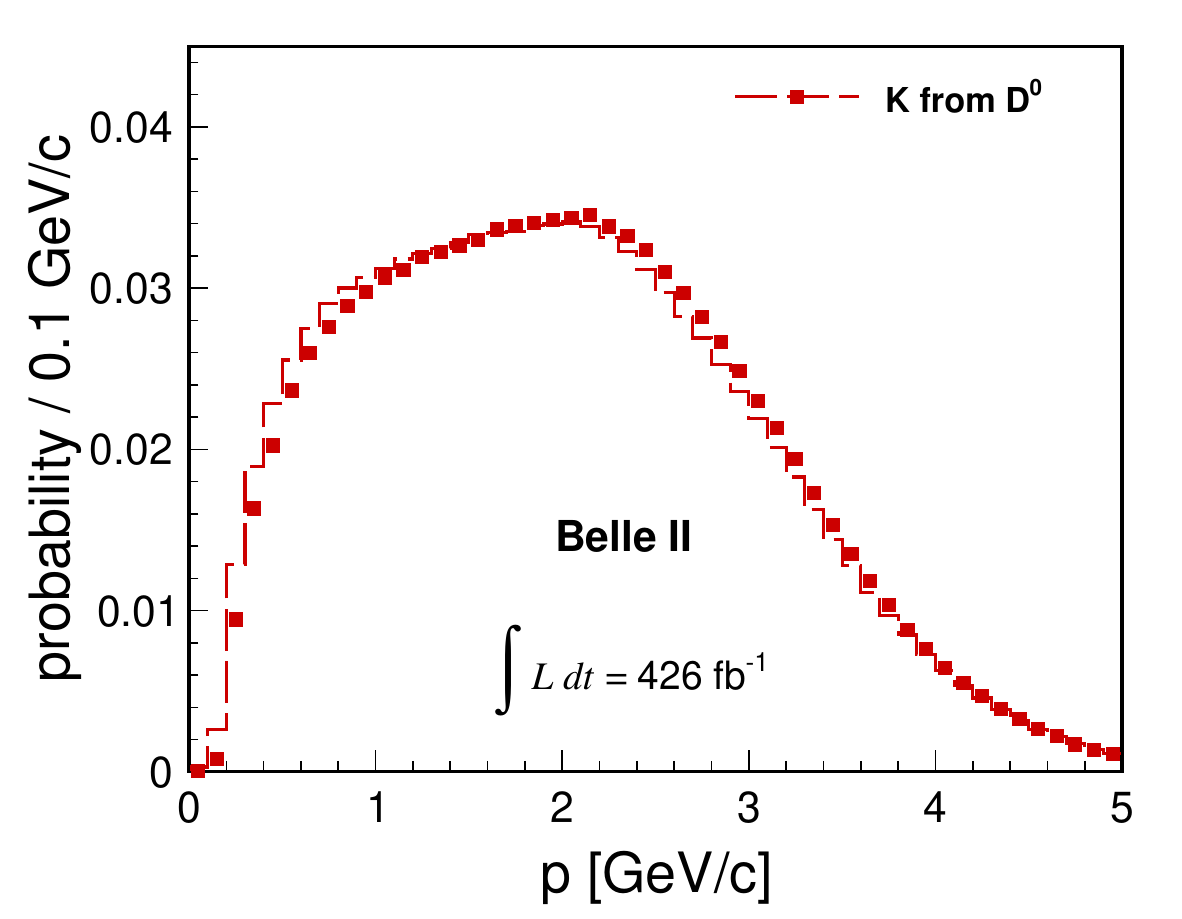} &
        \includegraphics[width=0.4\textwidth]{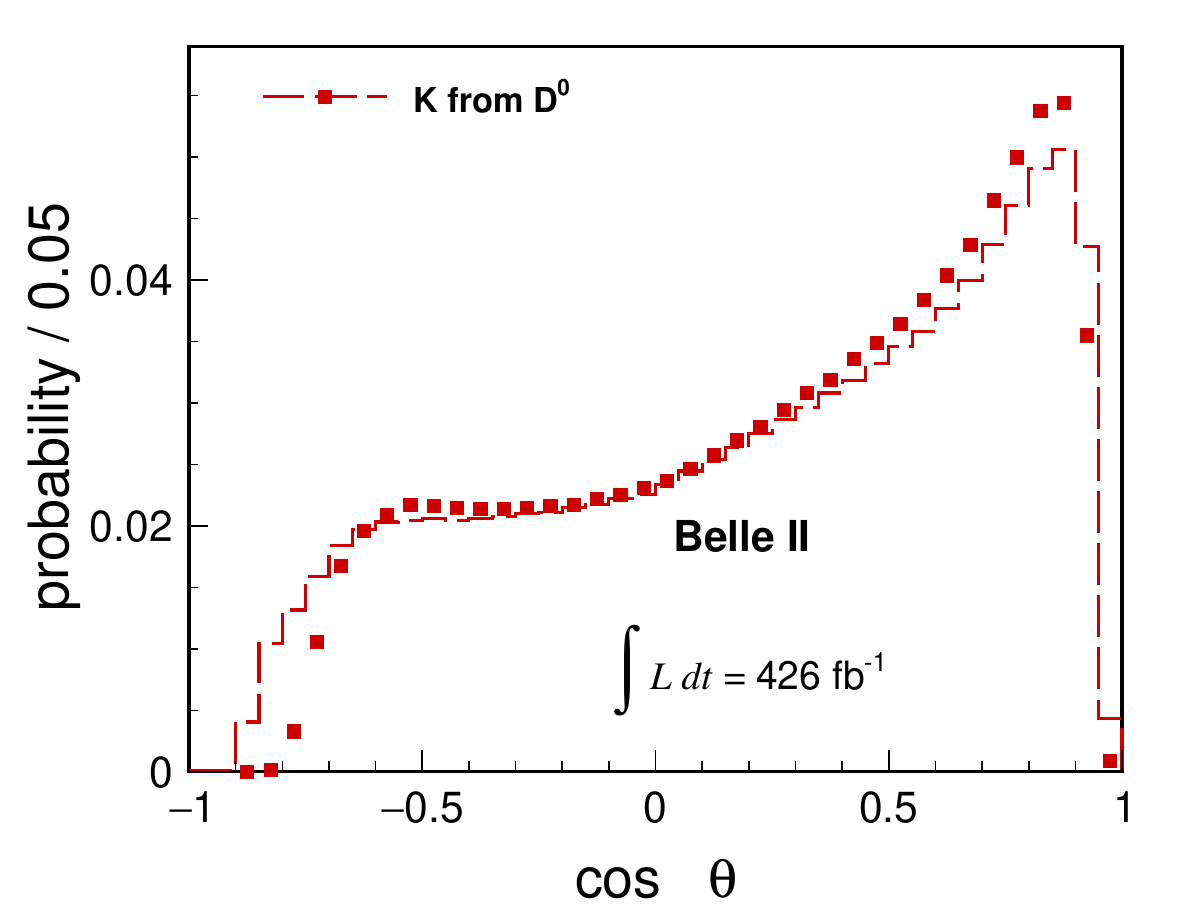} \\
        \includegraphics[width=0.4\textwidth]{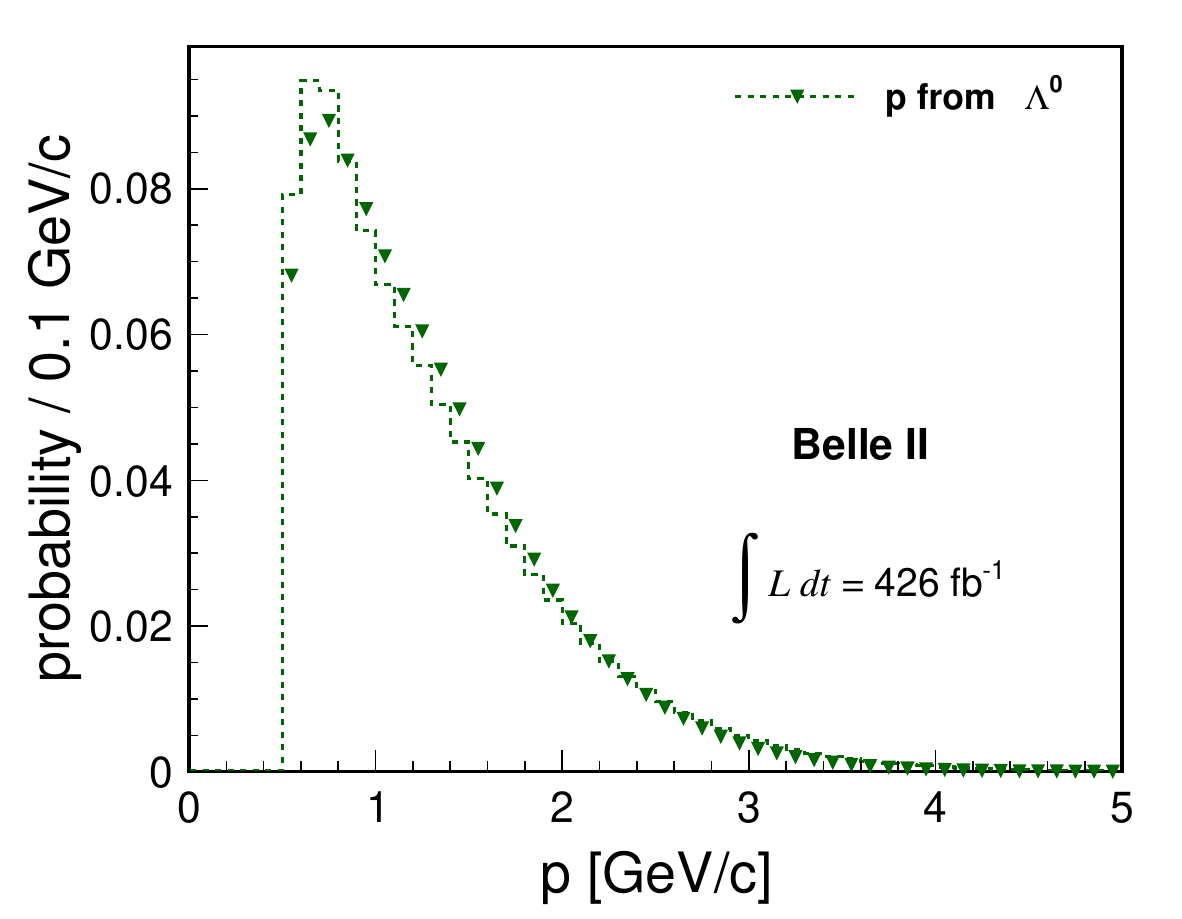} &
        \includegraphics[width=0.4\textwidth]{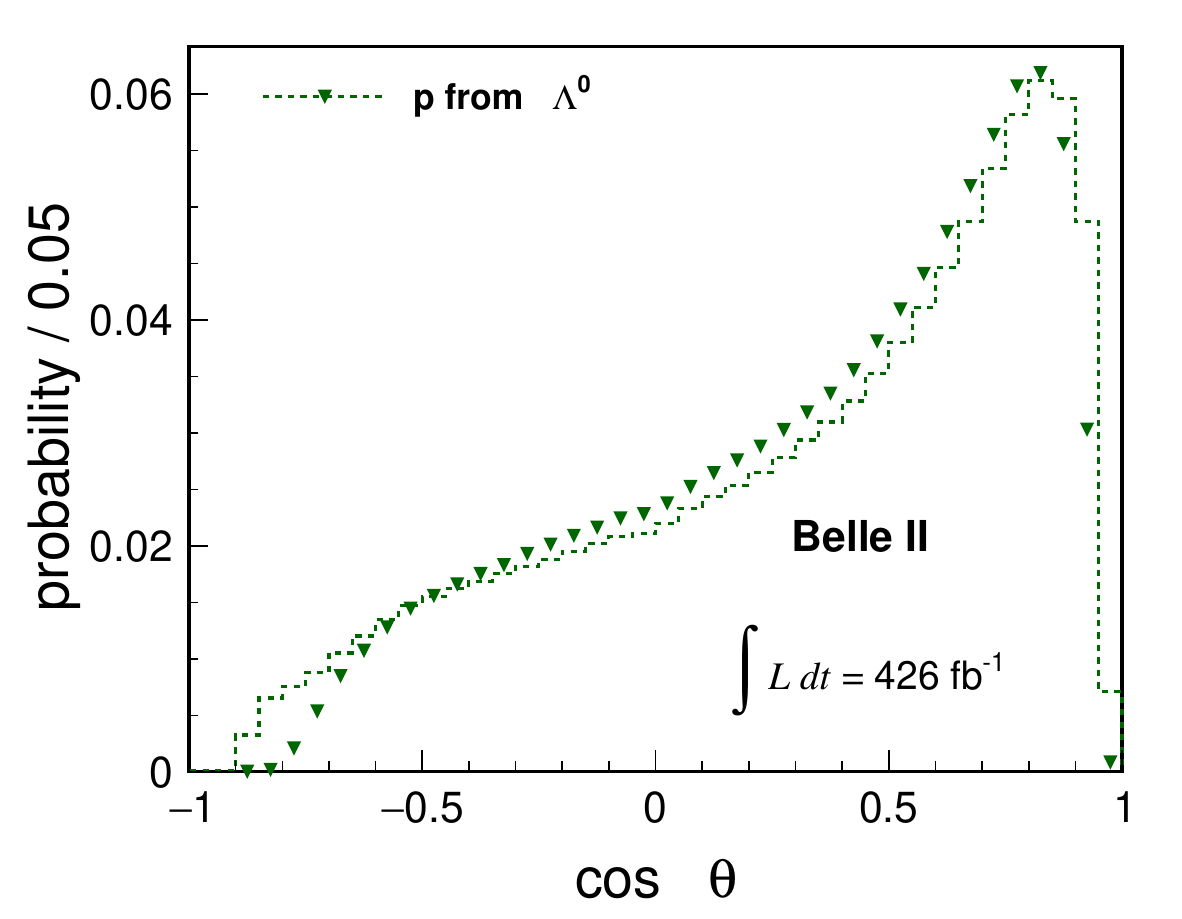} \\
    \end{tabular}   

        \caption{
            Background-subtracted $p$ and $\cos\theta$ distributions for pions, kaons, and protons from control samples in data~(points) and simulation~(lines). 
        }

        \label{fig:spectra_piKp}

    \end{center}
\end{figure*}

We fit to the mass distributions of reconstructed \PDzero, \PKzero, and \PLambda candidates to determine signal and background fractions and shapes, which allows us to determine the background-subtracted distributions of the particles in the control channels.
For \PDzero, we model signal as the sum of two Gaussian functions with a common mean; for the \PKzero, as the sum of a Gaussian function and a Johnson's $S_U$ function; and for \PLambda, as a Johnson's $S_U$ function.
For each channel, we model the background as a linear function.
In the regions of their nominal masses, the \PDzero, \PKzero, and \PLambda distributions have signal-to-background ratios of 28, 20, and 9, respectively.
Due to the higher signal-to-background ratio and broader momentum coverage, we normally use the \PDzero control sample for pions, while the \PKzero and \PLambda samples are used for cross-checks or for very low momenta.
Figure \ref{fig:spectra_piKp} shows the background-subtracted $\cos\theta$ and $p$ distributions for pions, kaons, and protons in the control channels.
They cover a wide momentum range and the entire detectable polar angle range ($\cos\theta \in [-0.86, 0.97]$); the small differences between data and simulation are acceptable, as our method to measure PID efficiencies does not require the simulation to perfectly reproduce the data.

We use four-lepton events, $\APelectron\Pelectron\to\APelectron\Pelectron\Pleptonplus\Pleptonminus$, to gather samples of electrons and muons to measure the probabilities that leptons are misidentified as hadrons.
The electron and positron from the colliding beams exchange virtual photons that produce a charged lepton pair that is detected, the beam electron and positrons, which are only slightly deflected by the interaction, continue down the beam pipe undetected.
The detected leptons have predominantly low momenta, allowing us to precisely determine lepton-as-hadron misID rates at low momenta, where they are highest.

We trigger the recording of such events with a purely track-based trigger, requiring the presence of two tracks that originate from the interaction region and are back-to-back in the transverse plane.
We do not trigger on ECL information because that could bias PID performance.
We select events with only two charged tracks, oppositely charged, each with momentum greater than \SI{0.4}{GeV/c} and point of closest approach to the interaction point less than \SI{5.0}{cm} in $z$ and \SI{2.0}{cm} in the transverse plane. To suppress cosmic muons reconstructed as two tracks, we require the opening angle (in three dimensions) of the tracks be less than \SI{168}{\degree}. In the c.m.\ frame, we require that the sum of the energies of charged and neutral particles detected in the event be less than \SI{6}{GeV} and the track pair have longitudinal momentum less than \SI{1.0}{GeV/c}, transverse momentum less than \SI{150}{MeV/c}, and mass less than \SI{3.0}{GeV/c^2}. Background events from $\APelectron\Pelectron\to\APelectron\Pelectron h^+ h^-$, $\APelectron\Pelectron\to h^+ h^- \gamma$ ($h = \pi, K$), and $\APelectron\Pelectron\to\tau^+\tau^-$ are subtracted using simulated events, corrected for the known discrepancies in angular distributions and hadron-ID performance. Their impact is further reduced by employing a tag-and-probe technique in which tight PID selection criteria are applied to only one of the candidate lepton tracks (the tag) while the other (the probe) is used to measure the PID performance without bias.
\section{Local likelihoods}
\label{sec:pid_model}

To identify particles, we use information from all subdetectors except the PXD.
For each subdetector, $d$, and particle species, $\alpha$, we define the local likelihood, 
$\mathcal{L}^d_\alpha(\vec{x}^{\,d} | \, \vec{p} \, )$, for the subdetector's measurements $\vec{x}^{\,d}$.

\subsection{SVD likelihoods}
\label{subsec:svd_likelihood}

\newcommand{\SVDdEdx}{\ensuremath{\eta}\xspace}

The SVD measures energy loss typically eight times for each track, twice in each of the four double-sided layers, from which we can calculate the specific energy loss, \dEdx*. To distinguish between particle species, the absolute calibration is not important, so we define a unitless variable proportional to \dEdx*, \SVDdEdx, as the energy loss divided by the average energy loss for an electron at the Fermi plateau~\cite{Chapter34_pdg}.

The species-dependent likelihood for all of a track's \SVDdEdx measurements is calculated using histogram templates of the two-dimensional $(p, \SVDdEdx)$ distributions, $\mathcal{H}\Sup{SVD}_\alpha$, determined from control channels in real and simulated data, 
\begin{equation}
  \mathcal{L}\Sup{SVD}_\alpha (\vec\SVDdEdx \, | \, p) \equiv \prod_{i} \mathcal{H}\Sup{SVD}_\alpha (\SVDdEdx_i | \, p),
\end{equation}
where the product runs over all the individual $\SVDdEdx$ measurements available for a track, excluding the two highest to reduce bias arising from the long tail of the Landau distribution.

\subsection{CDC likelihoods}
\label{subsec:cdc_likelihood}

\newcommand{\CDCdEdx}{\ensuremath{\bar\eta}\xspace}
\newcommand{\CDCdEdxMean}{\ensuremath{M_\alpha(p)}\xspace}
\newcommand{\CDCdEdxVariance}{\ensuremath{V(\CDCdEdx, \theta, n)}\xspace}

The CDC measures \dEdx* for each drift cell traversed by a track.
We average these measurements and normalize by the mean average for electrons at the Fermi plateau to calculate a unitless specific energy loss, \CDCdEdx.
When calculated with the lowest 5\% and highest 25\% of each track's measurements excluded, \CDCdEdx is Gaussian distributed, with mean, \CDCdEdxMean, that depends on $p$ and $\alpha$ and variance, \CDCdEdxVariance, that depends on \CDCdEdx, $\theta$, and the number of sense wires hit $n$.
The species-dependent likelihood for \CDCdEdx is therefore
\begin{equation}
    \mathcal{L}\Sup{CDC}_\alpha(\CDCdEdx, n | p, \theta)
    \equiv
    \exp[-\frac{[\CDCdEdx - \CDCdEdxMean]^2}{2 \CDCdEdxVariance}],
\end{equation}
where we neglect the species independent normalization of $1 / \sqrt{2\pi \CDCdEdxVariance}$.

\subsection{TOP likelihoods}
\label{subsec:top_likelihood}

The TOP's micro-channel-plate photomultiplier tubes have fine enough time resolution and short enough dead time to measure individual Cherenkov photons.
Given an incoming particle's momentum and species, we calculate the expected photon density in each channel, $c$, of the photon-detector array as a function of the time since the $e^+e^-$ collision, $S^c_\alpha(t|\vec{p})$. We account for geometric acceptance, scattering, absorption, and detection efficiency and model the photon density distribution as a sum of Gaussian functions, each weighted by its expected photon count~\cite{Staric:2011}.
The expected means, variances, and expected photon counts of the functions are analytically calculable from the particle's species and momentum.
The species-dependent likelihood for the photons measured is
\begin{multline}
    \mathcal{L}\Sup{TOP}_\alpha(\vec{c}, \vec{t}\,| \vec{p})
    \equiv \\ \mathcal{P}(N | \nu_\alpha(\vec{p})) \ \prod_{i=1}^N \frac{S^{c_i}_\alpha(t_i|\vec{p}) + B^{c_i}(t_i)}{\nu_\alpha(\vec{p})},
  \label{eqn:top-local-likelihood}
\end{multline}
where $c_i$ and $t_i$ are the channel and time in which photon $i$ is measured, $B^c(t)$ models background photons in channel $c$, and $\mathcal{P}$ is the Poisson probability to detect $N$ photons given we expect $\nu_\alpha(\vec{p})$, the sum of expected counts in all channels from both the particle and background, accounting for detection efficiencies.
In each channel, the expected count from the particle is the sum of photon-count weights and the expected count from background is the integral of $B^c(t)$ over the measurement time. We assume that the background is uniform in time and space and normalize $B^c(t)$ for each event by counting hits in TOP modules that are not traversed by any tracks.


\subsection{ARICH likelihood}
\label{subsec:arich_likelihood}

The ARICH's pixelated hybrid avalanche photodiodes record only binary signals: whether a pixel fired or not.
Given an incoming particle's momentum and species, we calculate the number of photons expected to hit each pixel, $\nu^c_\alpha(\vec{p})$. We account for geometric acceptance, scattering, absorption, beam backgrounds, and detection efficiency, and we add effective background photons to account for electronic noise and photodiode spuriously firing.
With $F$ the set of pixels that have fired in a particular event, the species-dependent likelihood for the states of the pixels is the product of Poisson probabilities for each unfired pixel, $c \notin F$, to have been hit by no photon, $\exp[-\nu^c_\alpha(\vec{p})]$, and for each fired pixel, $c \in F$, to have been hit by one or more photons, $1 - \exp[-\nu^c_\alpha(\vec{p})]$, given the expected number of photons in that pixel,
\begin{equation}
    \mathcal{L}\Sup{ARICH}_\alpha(\vec{p})
    \equiv \prod_{c\,\notin F} e^{-\nu^c_\alpha(\vec{p})} \times \prod_{c\,\in F} \qty[1 - e^{-\nu^c_\alpha(\vec{p})}].
\end{equation}
The products are calculated only over pixels within a ring on the photodiode plane defined by cones around $\vec{p}$ with opening angles \SI{0.1}{rad} and \SI{0.5}{rad}.

\subsection{ECL likelihood}
\label{subsec:ecl_likelihood}

The ECL measures the energy deposited by a particle in its crystals.
The ratio of this energy to $p$ strongly depends on the particle species.
The species-dependent likelihood for this ratio is calculated using Gaussian kernel-density-estimation templates~\cite{Cranmer:2000du} determined from simulated data of isotropically distributed particles.
They are determined separately for each region of the Cartesian product of three $\theta$ regions, three $p$ regions, and the two polarities of electric charge, $q$.
The boundaries of the $\theta$ regions are \SI{17}{\degree}, \SI{32}{\degree}, \SI{128}{\degree}, and \SI{150}{\degree}, which are the boundaries of the ECL's forward end cap, barrel, and backward end cap.
The boundaries of the two lower $p$ regions are \SI{0.2}{GeV/c}, \SI{0.6}{GeV/c}, and \SI{1.0}{GeV/c}; the third region contains all momenta greater than \SI{1.0}{GeV/c}.

\subsection{KLM likelihoods}
\label{subsec:klm_likelihood}

Muons with momenta greater than \SI{1}{GeV/c} typically traverse the whole KLM; hadrons typically stop in its first layers; and electrons rarely reach it. So we use the depth of hits in the KLM, along with information about their lateral shape, to distinguish between particle species.

We extrapolate each track from the CDC into the KLM using a Kalman filter, assuming it is a stable muon and accounting for multiple scattering and ionization energy loss.
Extrapolation halts when the track exits the KLM or the particle's energy falls below \SI{2}{MeV}.
We associate hits in the KLM to the track if the distance between the hit and the extrapolated track is less than 350\% of its uncertainty, which is due to uncertainties on the hit position and track extrapolation.


The species-dependent likelihood is the product of the likelihoods for the observed depth and the lateral shape, ignoring the potential correlation between the two.
From simulation, we compute the species-dependent probability that a track with extrapolated momentum $\vec{p}\sub{ex}$ reaches the $j^{\rm th}$ layer, $h_\alpha^j(\vec{p}\sub{ex})$.
The species-dependent likelihood for the depths of the hit layers, $j\in H$, and unhit layers, $j\notin H$ is therefore
\begin{multline}
  \mathcal{L}\Sup{KLM}_\alpha (H | \vec{p}\sub{ex})
  \equiv \\
  \prod_{j \in H} \epsilon^j\sub{KLM} h_\alpha^j(\vec{p}\sub{ex})
  \times 
  \prod_{j \notin H} \qty[1 - \epsilon^j\sub{KLM} h_\alpha^j(\vec{p}\sub{ex})],
\end{multline}
where $\epsilon^j\sub{KLM}$ is the efficiency to detect a particle in layer $j$ of the KLM.
Only layers through which the extrapolated track pass are included in the likelihood.
This set of layers is species independent since the track is extrapolated assuming it is a muon.
The species-dependent likelihoods for the lateral shape of the hits are template likelihoods for the $\chi^2$ of the Kalman filter, obtained from simulated data separately for each species.
\section{PID probabilities}
\label{sec:selectors}

Assuming the local likelihoods are well-formed and uncorrelated, we multiply them to form global species-dependent likelihoods
\begin{equation}
\mathcal{L}_\alpha \equiv \prod_d \mathcal{L}^d_\alpha.
\label{eqn:basic_pid_prob}
\end{equation}
Using Bayes' theorem and the law of total probability, assuming a uniform prior probability for all species, 
we compute the likelihood ratio
\begin{equation}
    P_{\alpha} = \frac{\mathcal{L}_\alpha}{\sum_\gamma \mathcal{L}_\gamma},
    \label{eqn:unweighted_pid_prob}
\end{equation}
where the sum in the denominator runs over all species~(\Pe, \Pmu, \Ppi, \PK, \Pp, \Pdeuteron).
In many analyses, we need only decide between a subset of possible species and restrict the values of considered species $\gamma$.
The binary likelihood ratio, restricting $\gamma$ to two species,
\begin{equation}
    P_{\alpha/\beta} \equiv \frac{\mathcal{L}_\alpha}{\mathcal{L}_\alpha + \mathcal{L}_\beta},
    \label{eqn:unweighted_bin_pid_prob}
\end{equation}
is especially useful. In the special case where all considered particle species are present with the 
same abundance, these likelihood ratios correspond to the probabilities that the particle
under study is of species $\alpha$.
Throughout the rest of the paper we will refer to these global and binary likelihood ratios 
as simple probabilities.

Most particles do not enter all subdetectors.
It is very rare that a particle enters both the TOP and the ARICH, and only particles with momenta greater than \SI{500}{MeV/c} reach the KLM.
When a particle does not enter a subdetector, we assign $\mathcal{L}^d_\alpha = 1$ for all $\alpha$.

These probabilities are easy to implement and maintain in \basfii. For example, we can easily exclude local likelihoods from subdetectors that were not well calibrated in some period. However, these probabilities have significant disadvantages that limit performance.
They neglect correlations, though we expect them, since the likelihoods all depend on the same track parameters.
They also do not account for inefficient or uninformative detectors, allowing probabilities to possibly favor one species over another based mostly on statistical fluctuations.
The probabilities also assume that the local likelihoods are well formed and normalized such that no species is favored over another. 

To quantify some of these effects, we replace $\mathcal{L}_\alpha$ in equation~(\ref{eqn:basic_pid_prob}) with 
\begin{equation}
\tilde{\mathcal{L}}_{\alpha} \equiv \exp(\sum_d w_{\alpha, d} \log \mathcal{L}^d_{\alpha}),
\end{equation}
thus allowing different weights for each local likelihood.
We use a sample of simulated data to optimize the weights to best identify pions, though other goals could be applied.
The optimal weights depend on the momentum and angle ranges of the training data, which can be chosen to suit each analysis's needs.

Table~\ref{tab:weight_matrix} shows an example of the matrix of weights obtained for particles (in the same amounts for all species) uniformly distributed in [0.5, 5.0] GeV/c and in the angular acceptance of the detector.
All CDC, ECL, and SVD likelihoods, except for the SVD's electron one, are weighted up;
all TOP, ARICH, and KLM likelihoods are weighted down.
These weights mostly account for the local likelihoods being poorly normalized.
For each detector, all weights are fairly similar.
Notably, the KLM's muon weight is more than twice the average of the KLM weights, demonstrating this subdetector's importance for identifying muons.

\begin{table*}[t]
    \caption{Local-likelihood weights of the reweighted global PID likelihood for particles uniformly distributed in [0.5, 5.0] GeV/c and the angular acceptance of the detector. 
    \label{tab:weight_matrix}}

    \newcommand{\wtcelltext}[2][]{\gradientcelld{#1}{#2}{0}{1}{3}{red}{white}{blue}{20}}
    \newcommand{\wtcell}[1]{\parbox{2.5em}{\raggedleft\wtcelltext{#1}{#1}}}


    \begin{tabular}{@{} p{3.5em} | *{5}{r}}
              & \Pe            & \Pmu           & \Ppi           & \PK            & \Pp            
              \\ \hline
        SVD   & \wtcell{0.81} & \wtcell{1.36} & \wtcell{1.06} & \wtcell{1.79} & \wtcell{1.72} \\ 
        CDC   & \wtcell{2.27} & \wtcell{1.96} & \wtcell{1.43} & \wtcell{1.91} & \wtcell{1.87} \\ 
        TOP   & \wtcell{0.43} & \wtcell{0.43} & \wtcell{0.43} & \wtcell{0.42} & \wtcell{0.41} \\ 
        ARICH & \wtcell{0.60} & \wtcell{0.61} & \wtcell{0.60} & \wtcell{0.61} & \wtcell{0.65} \\ 
        ECL   & \wtcell{2.46} & \wtcell{1.96} & \wtcell{2.02} & \wtcell{1.66} & \wtcell{1.96} \\ 
        KLM   & \wtcell{0.16} & \wtcell{0.48} & \wtcell{0.18} & \wtcell{0.22} & \wtcell{0.22} \\ 

    \end{tabular}

    \begin{center}
        \newcounter{wtsNcells} 
        \setcounter{wtsNcells}{16}
        \newcounter{wtscellnumber}
        \setlength{\tabcolsep}{0pt}
        \begin{tabular}{r *{\value{wtsNcells}}{p{1em}} l}
            0 \ \forloop{wtscellnumber}{0}{\value{wtscellnumber} < \value{wtsNcells}}{& \wtcelltext{\fpeval{3/\value{wtsNcells} * \value{wtscellnumber}}}} & \ 3
        \end{tabular}
    \end{center}
    
\end{table*}

\subsection{Neural-network \texorpdfstring{\PK-\Ppi}{kaon-pion} separation}

We can improve PID performance by combining the log likelihoods in a way that accounts for performance changing with a particle's momentum, direction, and charge.
We use a neural network to combine the thirty-six $\log\mathcal{L}^d_\alpha$ and the particle's $p$, $\theta$, $\phi$, and $q$ to replace $\mathcal{L}_{\Ppi}$ and $\mathcal{L}_{\PK}$ in equation~(\ref{eqn:unweighted_bin_pid_prob}).

We train the network using a sample of simulated particles containing equal amounts of pions and kaons, both positive and negative, isotropically distributed and uniformly distributed in momentum in the laboratory frame.
Training on this balanced sample avoids creating a network that distinguishes species based on kinematic variables instead of PID likelihoods.
We use this sample instead of a control channel in real data in order to cover the full momentum and angle ranges.

We train the network using the Adam optimizer~\cite{Kingma:2014vow} with two fully connected
hidden layers and use the training epoch with the largest area under its receiver-operating-characteristic curve, calculated using an independent simulated data set.
\section{PID efficiency and misID rate}
\label{sec:pid_efficiency_and_misID_rate}

Most \BelleII analyses estimate systematic uncertainties that arise from correcting PID efficiencies estimated from simulation so that they match the real efficiencies in data.
We use the control channels to measure PID efficiencies in data and calculate scaling factors to correct the simulation-based estimations.

To calculate the efficiency for identifying species $\alpha$, we inspect tracks that are identified as that species in the control channels.
From the fits to the mass distributions described in Section~\ref{sec:data} (without PID criteria), we calculate sWeights~\cite{Pivk:2004ty} for control-channel events that are uncorrelated with the quantities on which PID relies.
For a specific PID criterion, we define the $\alpha$-ID efficiency to be the sum of sWeights for events wherein the relevant track passes the criterion divided by the sum of sWeights of all events considered. Its statistical uncertainty is computed accounting for the individual weights of each event. 

Similarly, we calculate the probability to misID a particle of species $\beta$ as species $\alpha$ by applying the $\alpha$-ID criterion to a track assumed to be of species $\beta$ in a control channel that gives us such a sample.
The $\beta$-as-$\alpha$ misID rate is the sum of sWeights of events in which the track passes the $\alpha$-ID criterion divided by the sum of the sWeights of all the events considered.

We calculate these efficiencies and misID rates in subregions of momentum and polar angle.
The efficiency-correction factor is defined as the efficiency calculated using data divided by the efficiency calculated from simulation.
Statistical uncertainties on the efficiencies and systematic uncertainties from the mass fits are propagated to uncertainties on these correction factors.
These uncertainties are treated as systematic uncertainties from particle identification in \BelleII analyses.

We automate the above calculations using a systematic-corrections framework inspired by Ref.~\cite{Anderlini:2016kco} and written specifically for \basfii.
It manages simulation of the control channels, the selection of their samples in simulated and real data, and the fits to the mass distributions and subsequent sWeight calculation, using \BelleII's \texttt{b2luigi} workflow management system~\cite{b2luigi:doc}.
The framework greatly simplifies the calculation of analysis-specific correction factors and their related systematic uncertainties.
For each \BelleII analysis, one can calculate the systematic uncertainties from PID and other important sources such as tracking or \Ppizero, \PKshortzero, \PKlongzero, or \PLambda reconstruction.
\section{Hadron-identification performance}
\label{sec:performance}

We often only need to distinguish between two possible particle species.
We evaluate our abilities to separate kaons from pions, comparing the simple and neural-network PID probabilities, and to separate protons from kaons, using only the simple PID probability.
We also evaluate our lepton-as-pion misID rates. Hadron-ID performance was stable over the full data-taking period, except for effects from beam-induced backgrounds, so we use the full data set in all the following results.

\subsection{\texorpdfstring{\PK-\Ppi}{kaon-pion} separation}

We use the \PDzero control sample to evaluate kaon-pion separation. 
To evaluate the performance of each subdetector, we use the local-likelihood ratio,
\begin{equation}
    P^d_{\alpha/\beta} \equiv \frac{\mathcal{L}^d_\alpha}{\mathcal{L}^d_{\alpha} + \mathcal{L}^d_\beta},
\end{equation}
which, using Bayes' theorem and the law of total probability, assuming equal \textit{a priori} probabilities for $\alpha$ and $\beta$, is the local binary probability.

\begin{figure*}[t!]
  \begin{center}

      \includegraphics[width=\twocolumnplotwidth]{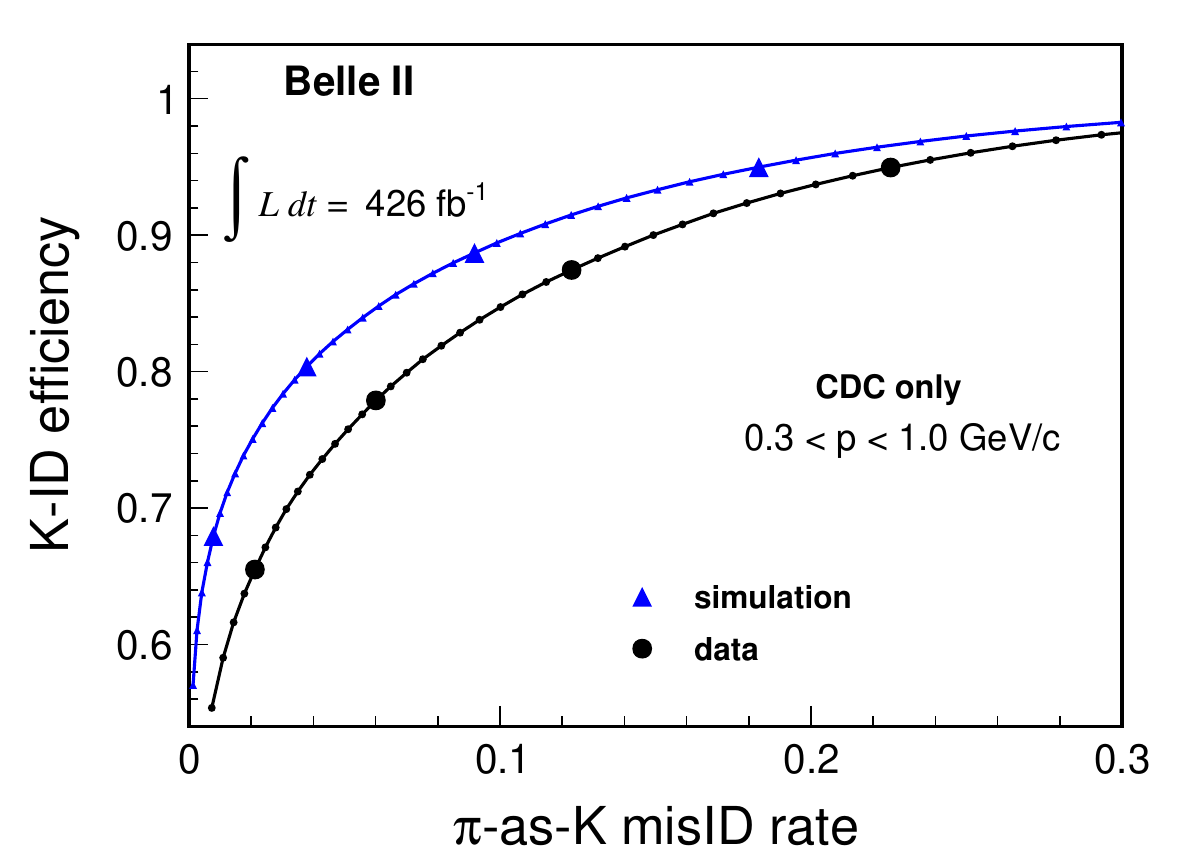}  \hfill%
      \includegraphics[width=\twocolumnplotwidth]{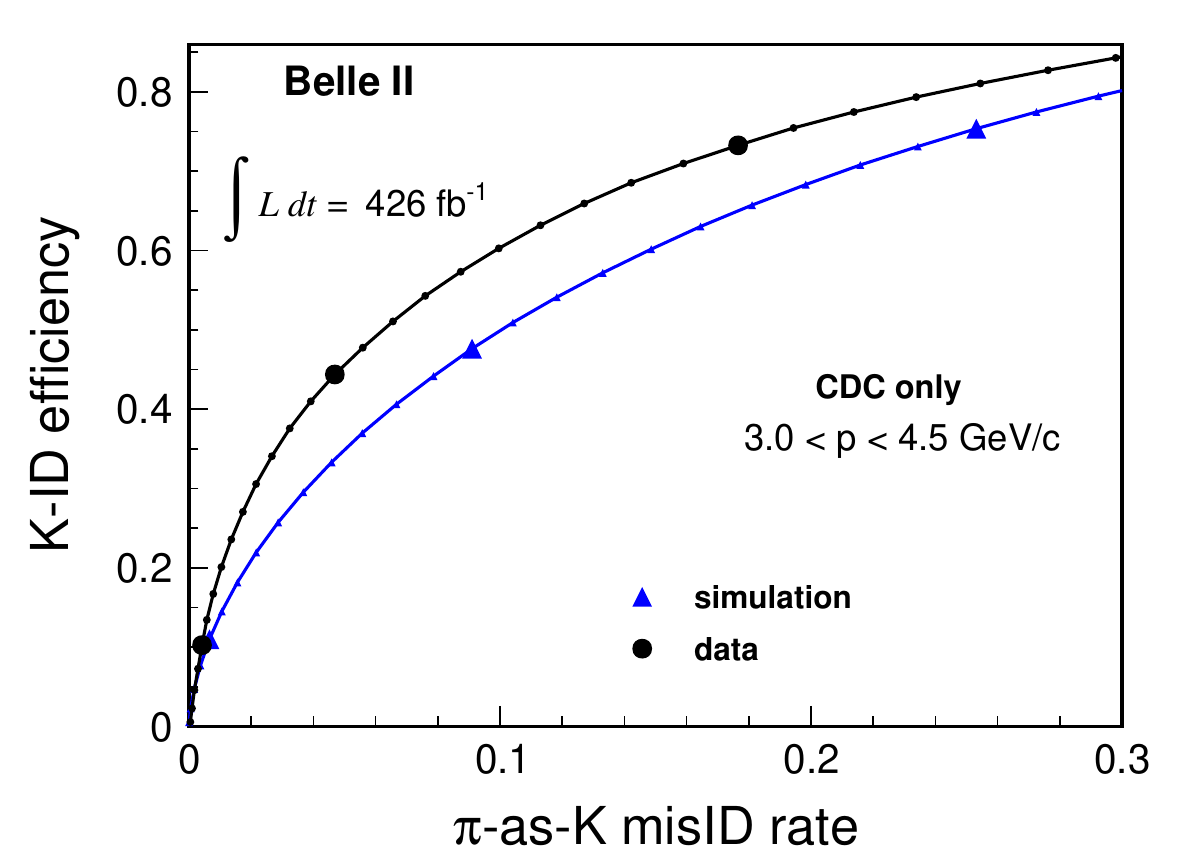} \\
      \includegraphics[width=\twocolumnplotwidth]{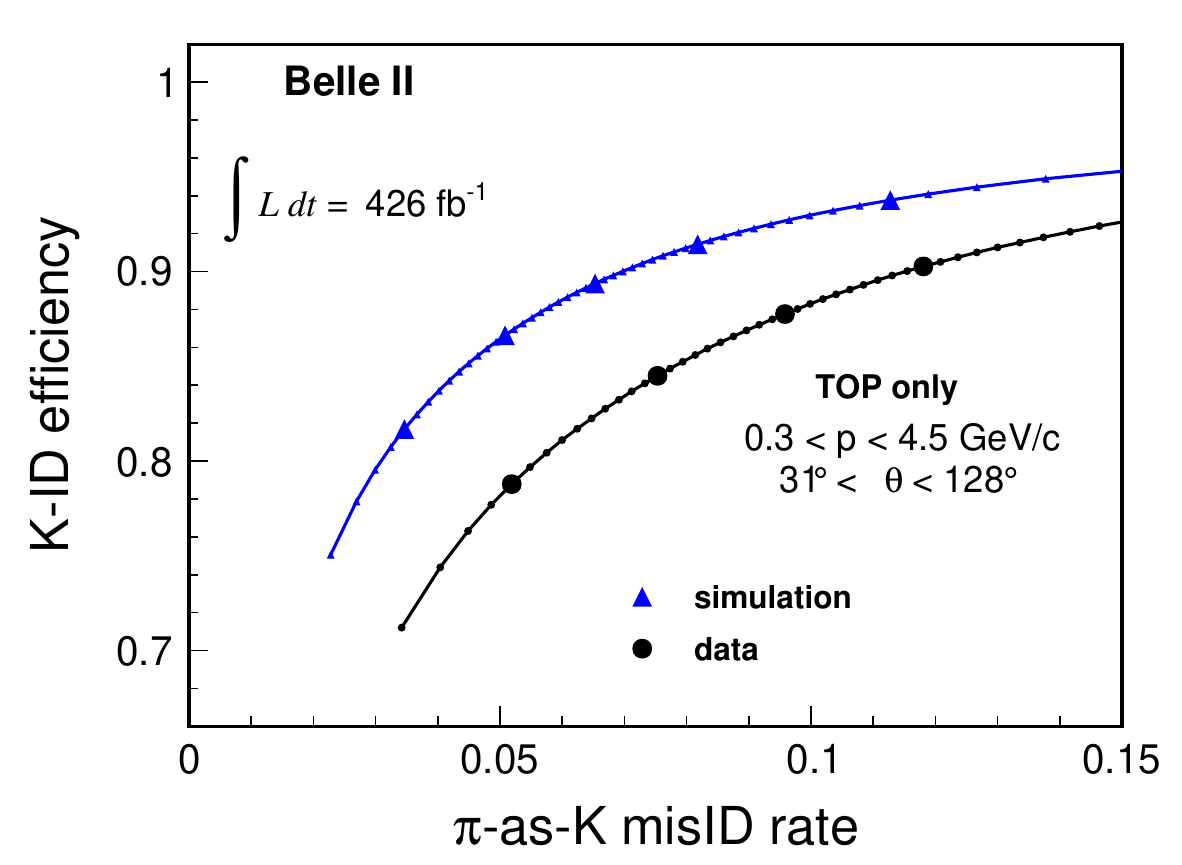}       \hfill %
      \includegraphics[width=\twocolumnplotwidth]{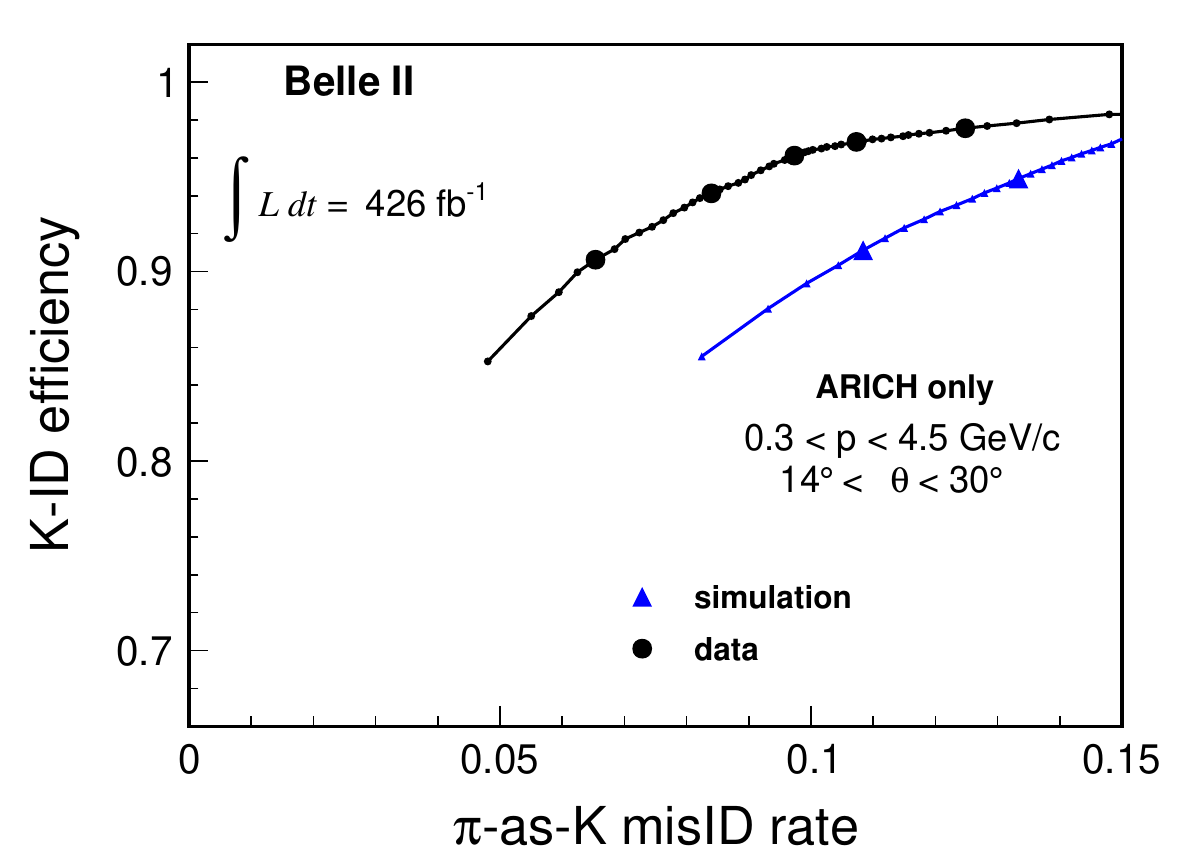}     \\

    \caption{
        \PK-ID efficiency as functions of the \Ppi-as-\PK misID rate for the local binary probabilities from the CDC~(for low and high momenta), TOP, and ARICH.
        The large markers show the results for the thresholds (from left to right in the plots) 0.9, 0.7, 0.5, 0.3, and 0.1; the lower three thresholds are not always visible in the ranges shown.
    }

    \label{fig:eff-misID_local-likelihood_kpi}

  \end{center}
\end{figure*}

Figure~\ref{fig:eff-misID_local-likelihood_kpi} shows the \PK-ID efficiency as functions of the \Ppi-as-\PK misID rate for PID requirements using only $P^d_{\PK/\Ppi}$ for $d = $ CDC, TOP, ARICH---the three subdetectors most important for \PK-\Ppi separation.
In each plot, the points for each data sample show the efficiency and the misID rate for the same set of $P^d_{\PK/\Ppi}$ thresholds, from zero~(at high \PK-ID efficiency) to 0.98~(at low \PK-ID efficiency) in steps of 0.02.

For the CDC, we show results separately for low and high momenta, excluding $p\in [1.0, 3.0]$ GeV/c since kaons and pions in this range have similar \dEdx* in the CDC.
For the same $P\Sup{CDC}_{\PK/\Ppi}$ requirement, the ID efficiency is always higher and the misID rate is always lower in the simulation than in data at low momentum; the opposite is true at high momentum.
This difference originates mostly from miscalibration of the simulation and it tends to cancel in samples with a broad momentum distribution extending above and below the crossing point of the \Ppi and \PK \dEdx* bands.
The Belle~II CDC \dEdx* simulation operates at the track-level: \dEdx*
values are randomly generated using the same parameterizations of mean and width that are used to calculate a $\chi^2$ value for a given hypothesis. This method
relies on high-quality data calibration, while imperfections in the data will generally lead to data-MC disagreement.  
An example is given by residual dependencies of the mean on polar angle, or a lack
of ``$\beta\gamma$ universality" (i.e., the corrected \dEdx* should depend 
only on $\beta\gamma = p/m$).  

For the TOP and ARICH, we show results only for particles that enter them.
The TOP performs worse in real data than in simulation.
This is due to instrumental effects that are not well simulated and from the simplified modeling of $B^c(t)$ in equation~\ref{eqn:top-local-likelihood}.
The ARICH performs better in real data than in simulation.
This difference originates from a fault in its digitization software that affects the simulation and that will be corrected in the next cycle of data reprocessing, improving the performance in simulation and making it more similar to the data.

\begin{figure*}[t!]
    \begin{center}
  
        \includegraphics[width=\twocolumnplotwidth]{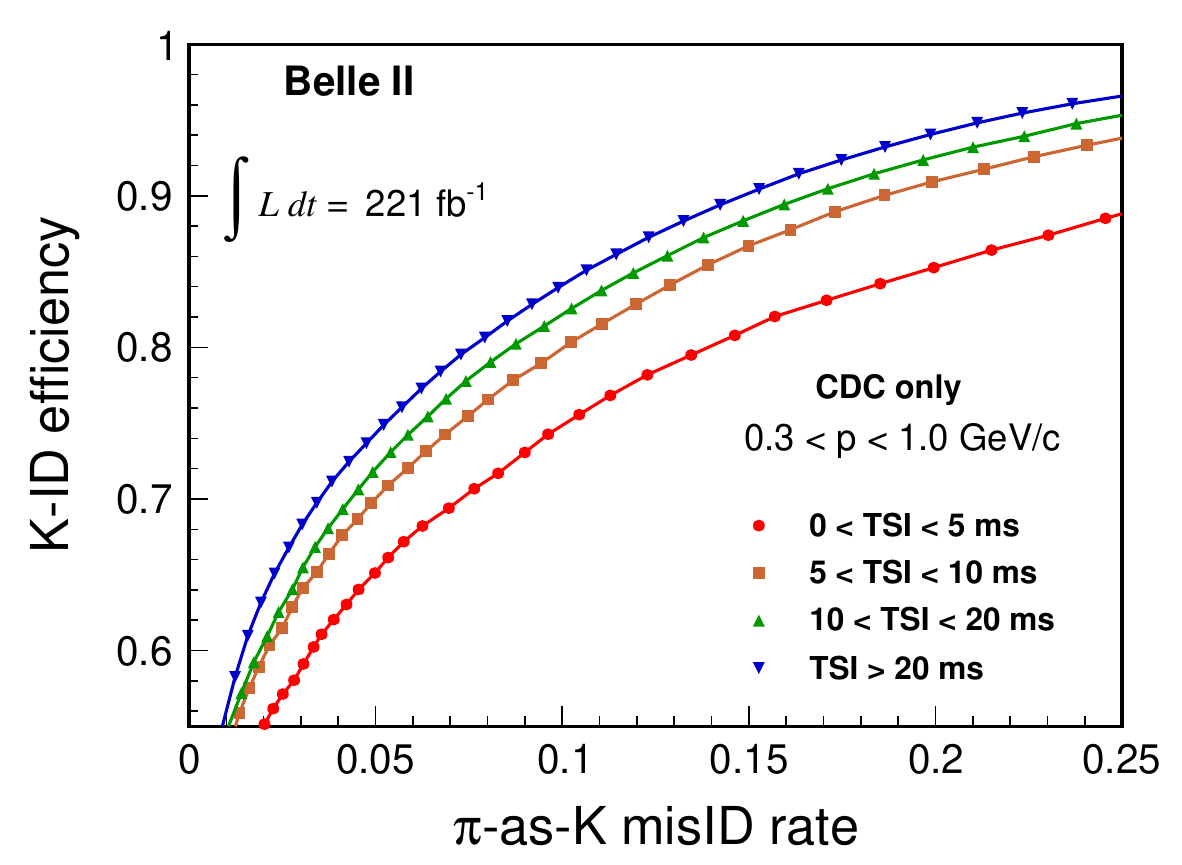}%
        \hfill%
        \includegraphics[width=\twocolumnplotwidth]{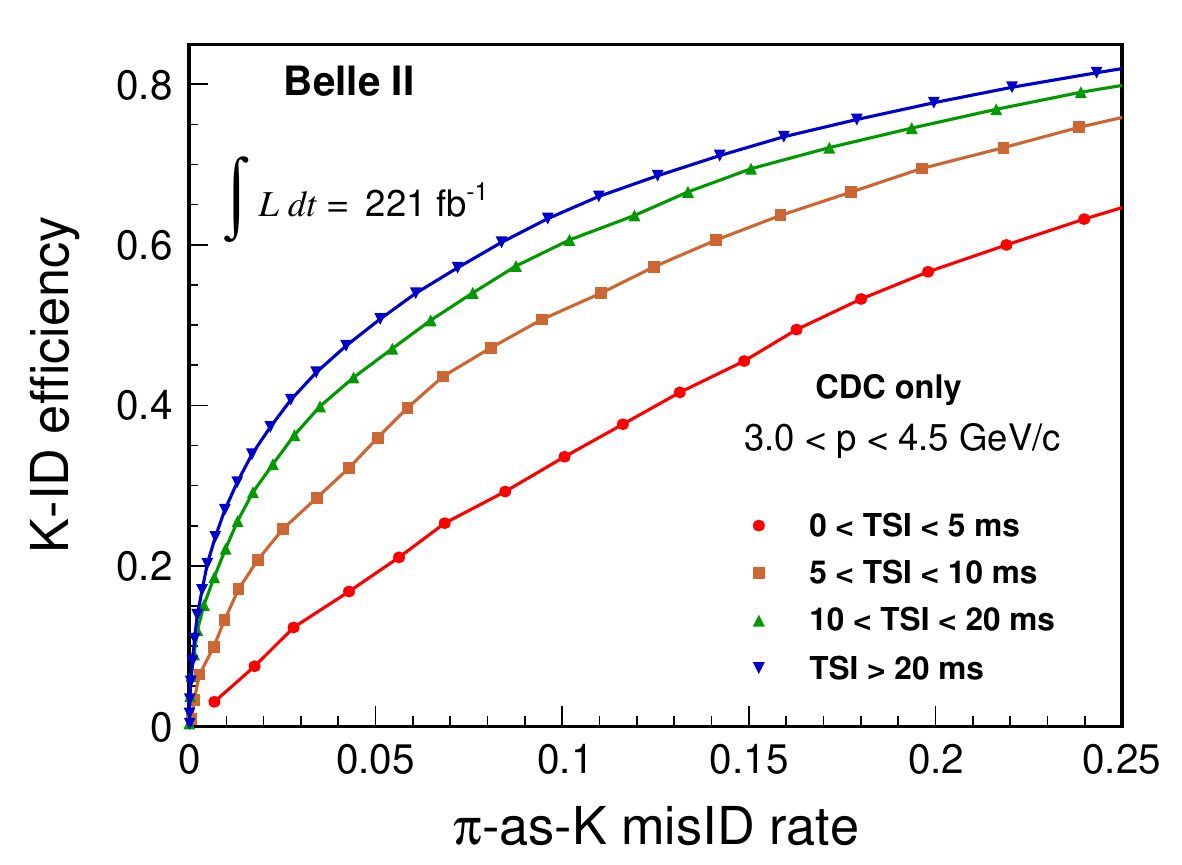}\\

        \caption{
            \PK-ID efficiency as functions of the \Ppi-as-\PK misID rate for the local binary probabilities from the CDC for low momenta~(left) and high momenta~(right) for four ranges of the time since beam injection.
        }
      
        \label{fig:CDC_ROCs_TSIbins}
    
  \end{center}
\end{figure*}

There are biases in the CDC's measurement of \dEdx* that are due to backgrounds related to beam injection that are not modeled in the simulation.
These biases are prominent in data collected between the end of 2021 and the middle of 2022, when the backgrounds originating from freshly injected \APelectron or \Pelectron bunches were particularly high.
Figure~\ref{fig:CDC_ROCs_TSIbins} shows the \PK-ID efficiency as functions of the \Ppi-as-\PK misID rate separately for low and high momenta in data collected at four different ranges of the time since the last \APelectron- or \Pelectron-beam injection~(TSI):
[0, 5] ms, containing 7\% of events;
[5, 10] ms, containing 10\%;
[10, 20] ms, containing 21\%;
and above 20 ms, containing 62\%.
In both momentum ranges, the \Ppi-as-\PK misID rate decreases with increasing TSI. 
The \PK-\Ppi separation performance improves with increasing TSI---that is, as the beam-injection backgrounds fall.
The effect is more pronounced at high momenta.

\begin{figure}[t!]
    \begin{center}

        \includegraphics[width=\textwidth]{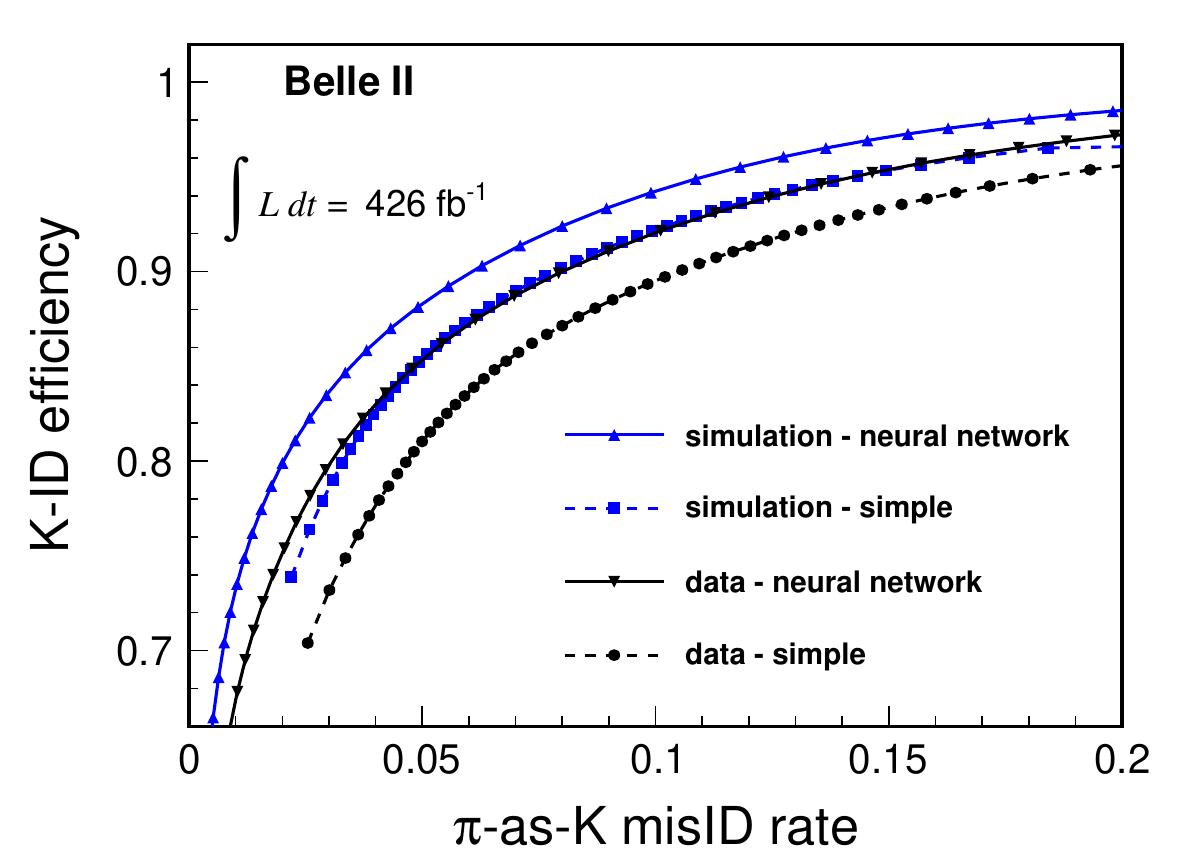}
    
        \caption{
            \PK-ID efficiency as functions of the \Ppi-as-\PK misID rate for the simple and neural-network \PK-\Ppi binary probabilities for data and simulation.
        }
    
        \label{fig:eff-misID_NN_kpi}
  
    \end{center}
\end{figure}

Figure~\ref{fig:eff-misID_NN_kpi} shows the efficiency as functions of the misID rate for PID requirements using the simple and neural-network \PK-\Ppi binary probabilities with information from all subdetectors in real and simulated data.
The performance in real and simulated data agree better when all subdetectors are used.
\PK-\Ppi separation is better with the neural network.

\begin{figure*}[t!]
    \begin{center}

        \includegraphics[width=\twocolumnplotwidth]{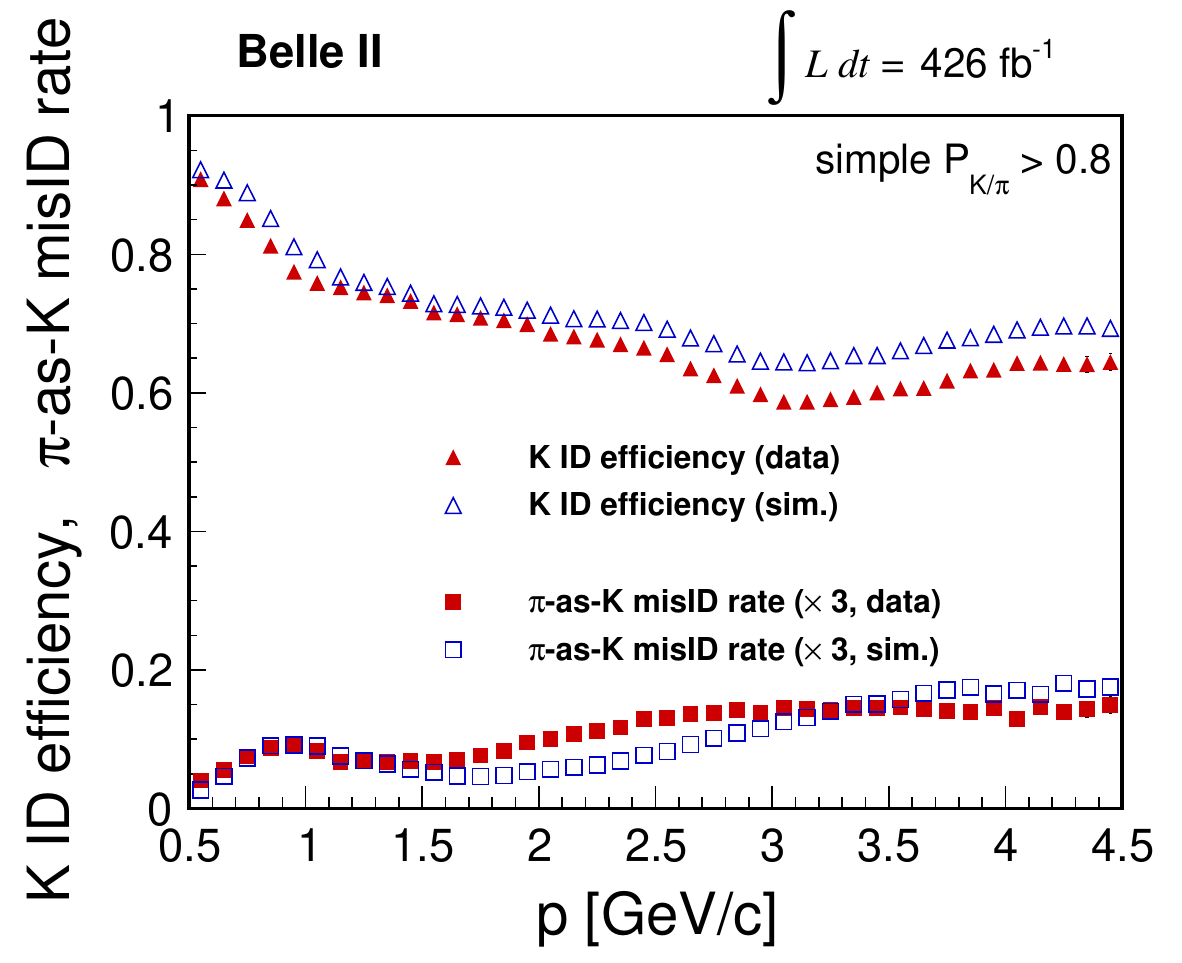}     \hfill%
        \includegraphics[width=\twocolumnplotwidth]{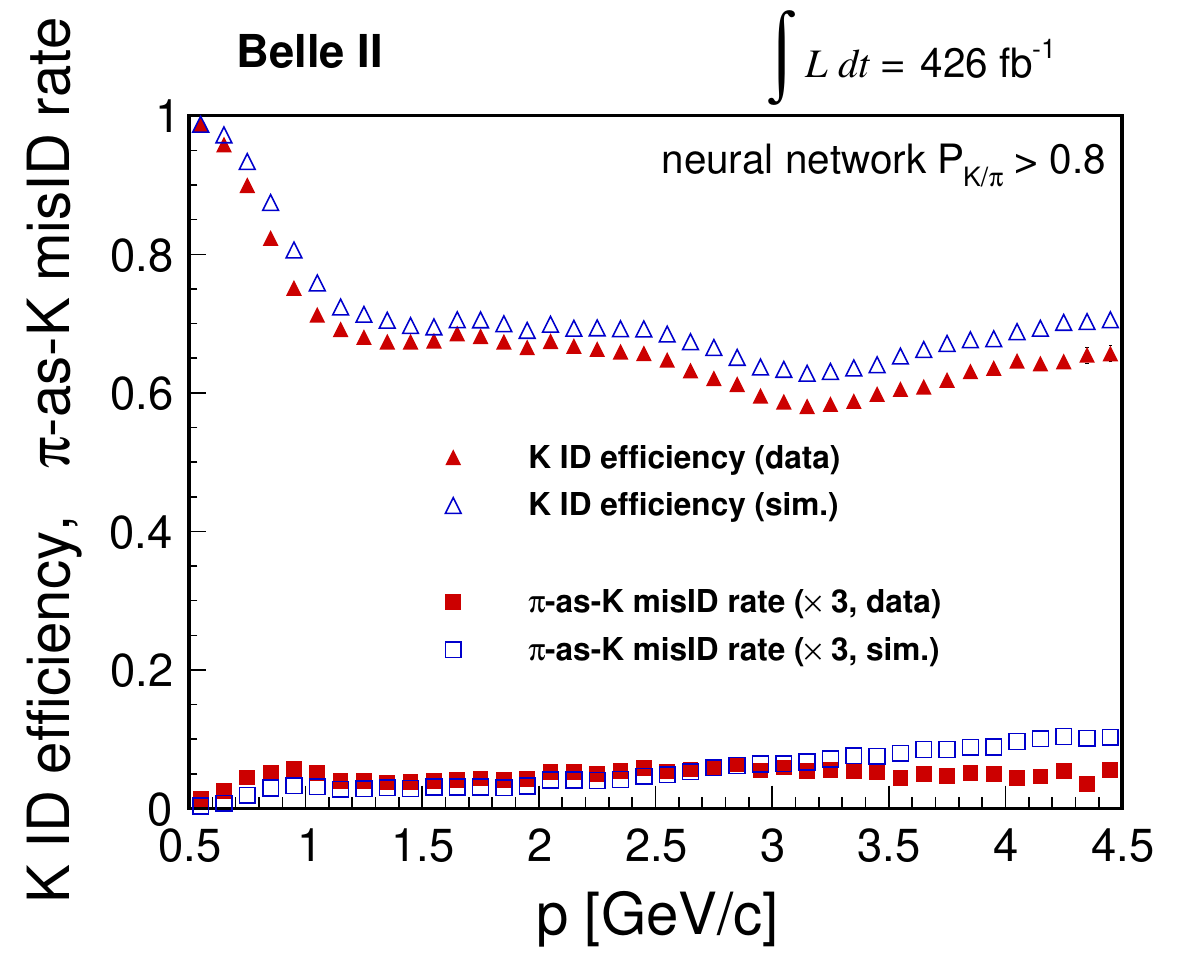}   \\
        \includegraphics[width=\twocolumnplotwidth]{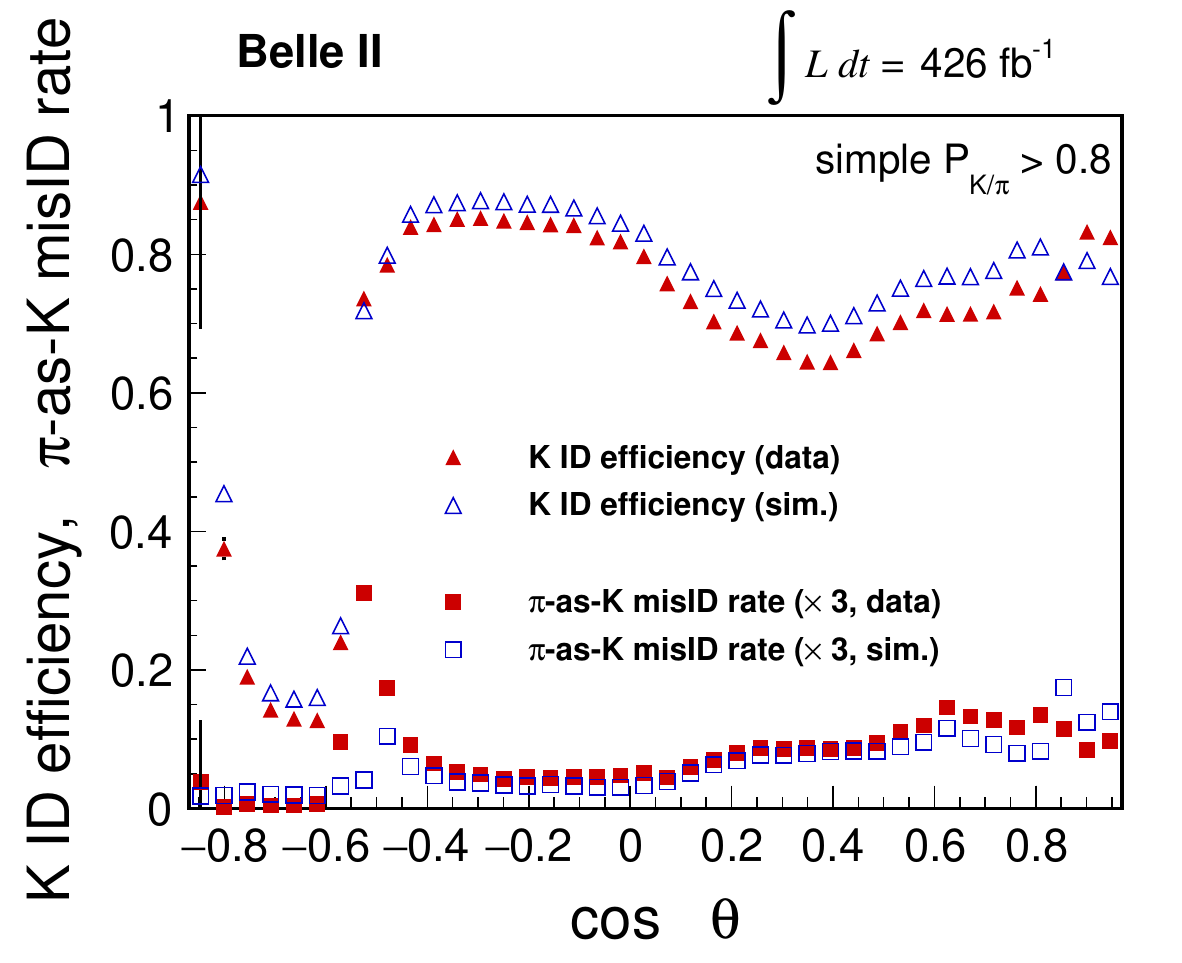}   \hfill%
        \includegraphics[width=\twocolumnplotwidth]{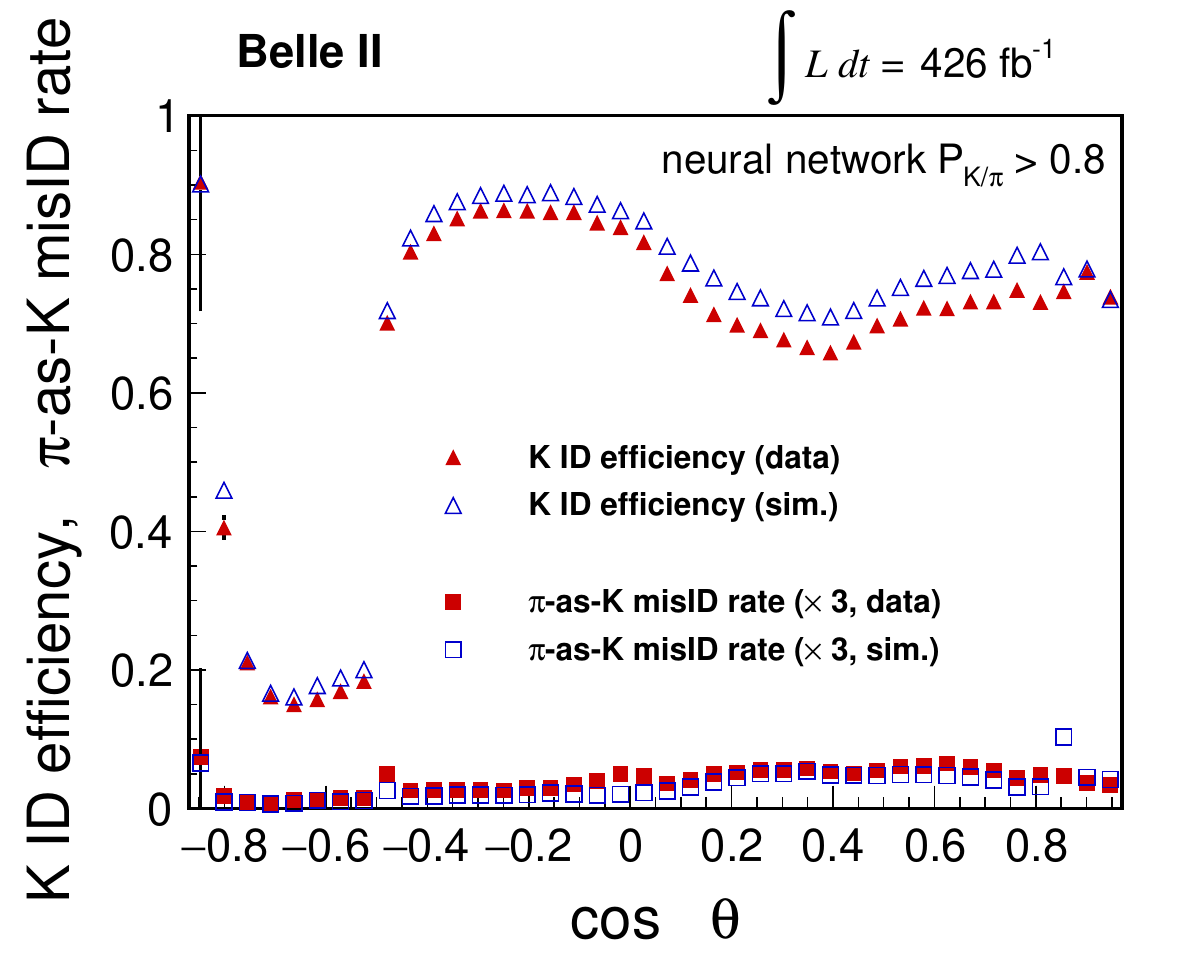} \\

        \caption{
            \PK-ID efficiency and \Ppi-as-\PK misID rate as functions of $p$ and $\cos\theta$ for $P_{\PK/\Ppi} > 0.8$ using the simple~(left) and neural-network~(right) probabilities. To aid visibility, the \Ppi-as-\PK misID rates are scaled by a factor of 3.
        }
        
        \label{fig:eff-misID_pcosth_kpi}
    
    \end{center}
\end{figure*}

Figure~\ref{fig:eff-misID_pcosth_kpi} shows the efficiency and misID rate as functions of $p$ and $\cos\theta$ for $P_{\PK/\Ppi} > 0.8$ for both the simple and neural-network binary probabilities in data and simulation.
Neural-network \PK-\Ppi separation is consistently more efficient with lower misID rates everywhere in the ranges of both $p$ and $\cos\theta$, in both data and simulation.
The \PK-\Ppi separation changes abruptly at $\cos\theta = -0.5$, which corresponds to the backward edge of the acceptance of the TOP detector.

\begin{figure*}[t!]
    \begin{center}

        \includegraphics[width=\twocolumnplotwidth]{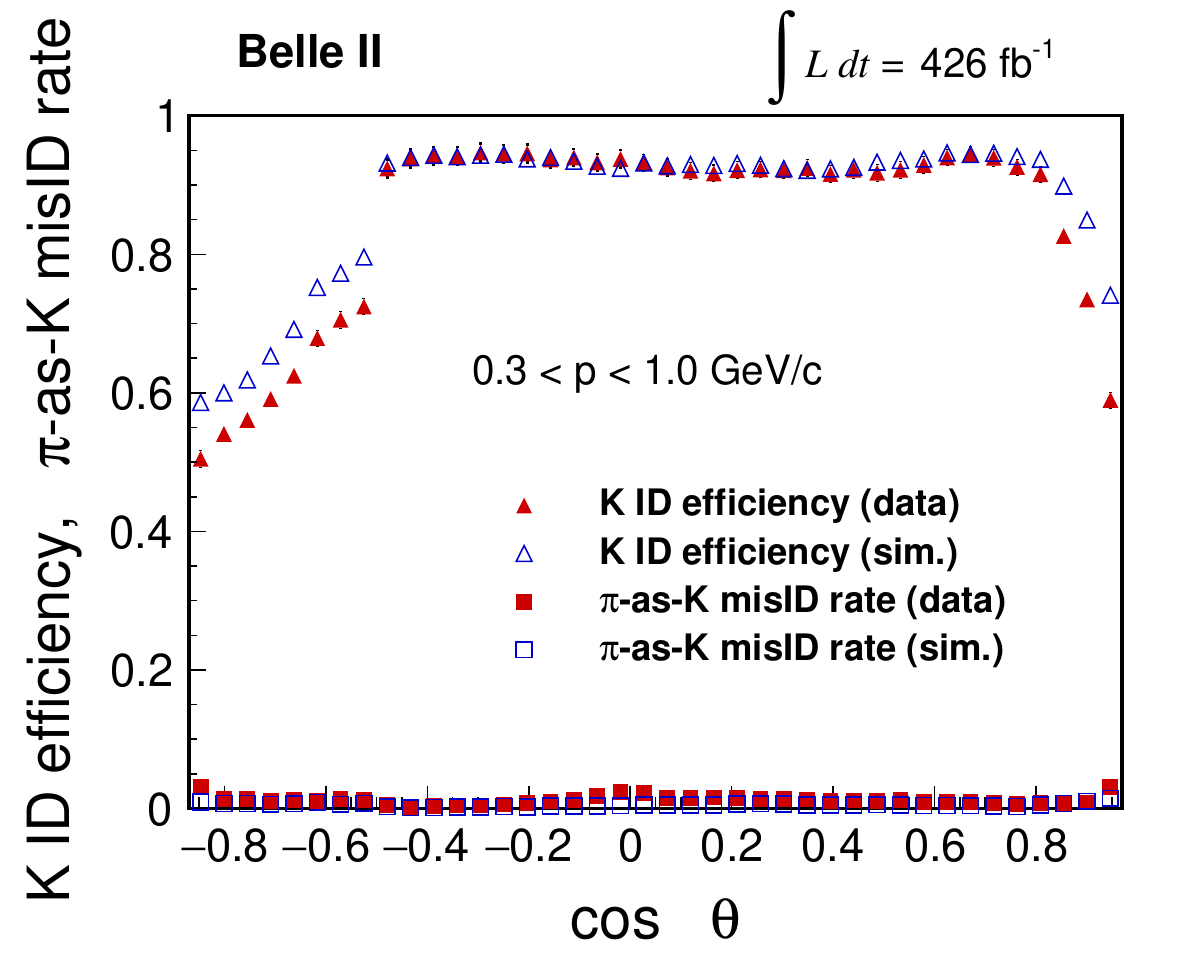}     \hfill%
        \includegraphics[width=\twocolumnplotwidth]{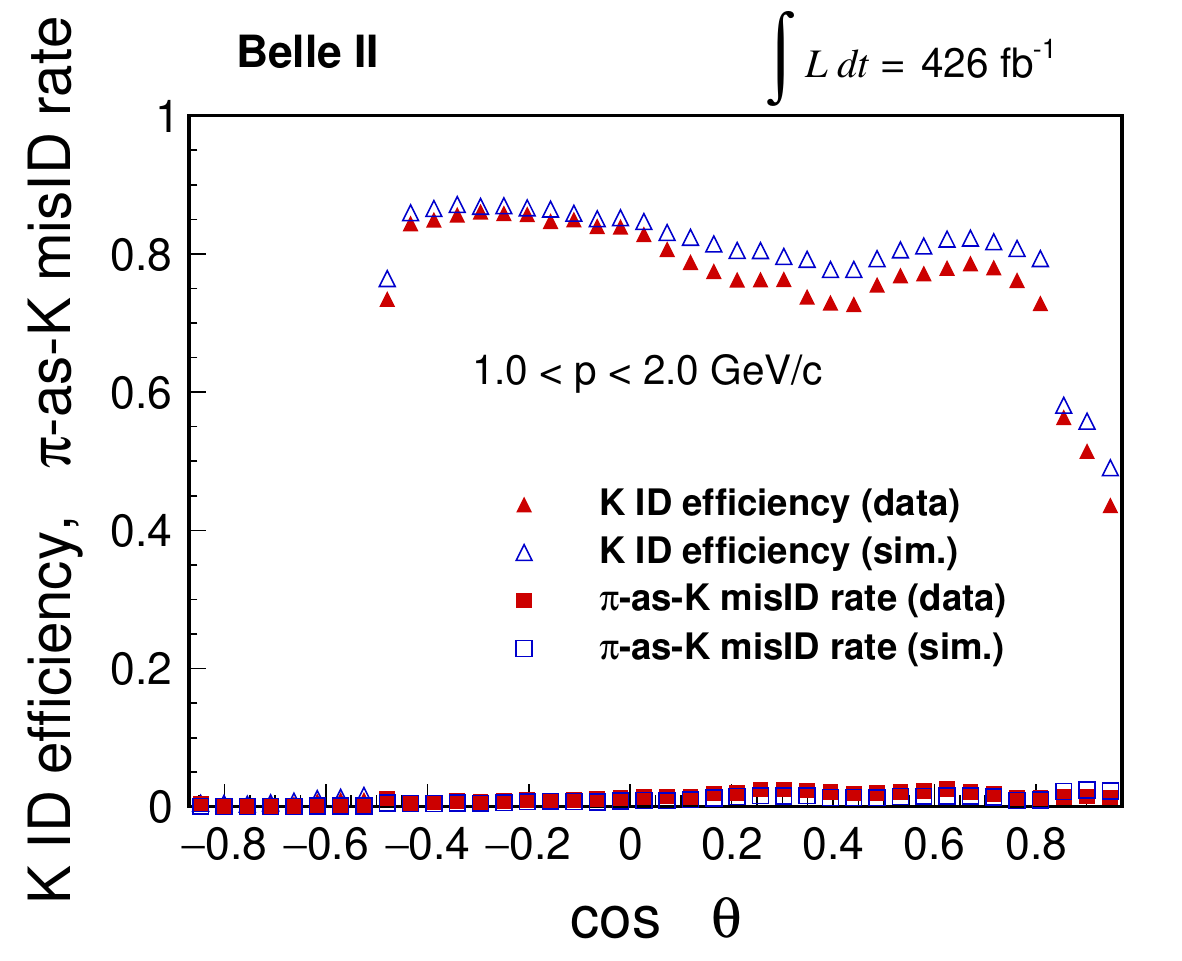}   \\
        \includegraphics[width=\twocolumnplotwidth]{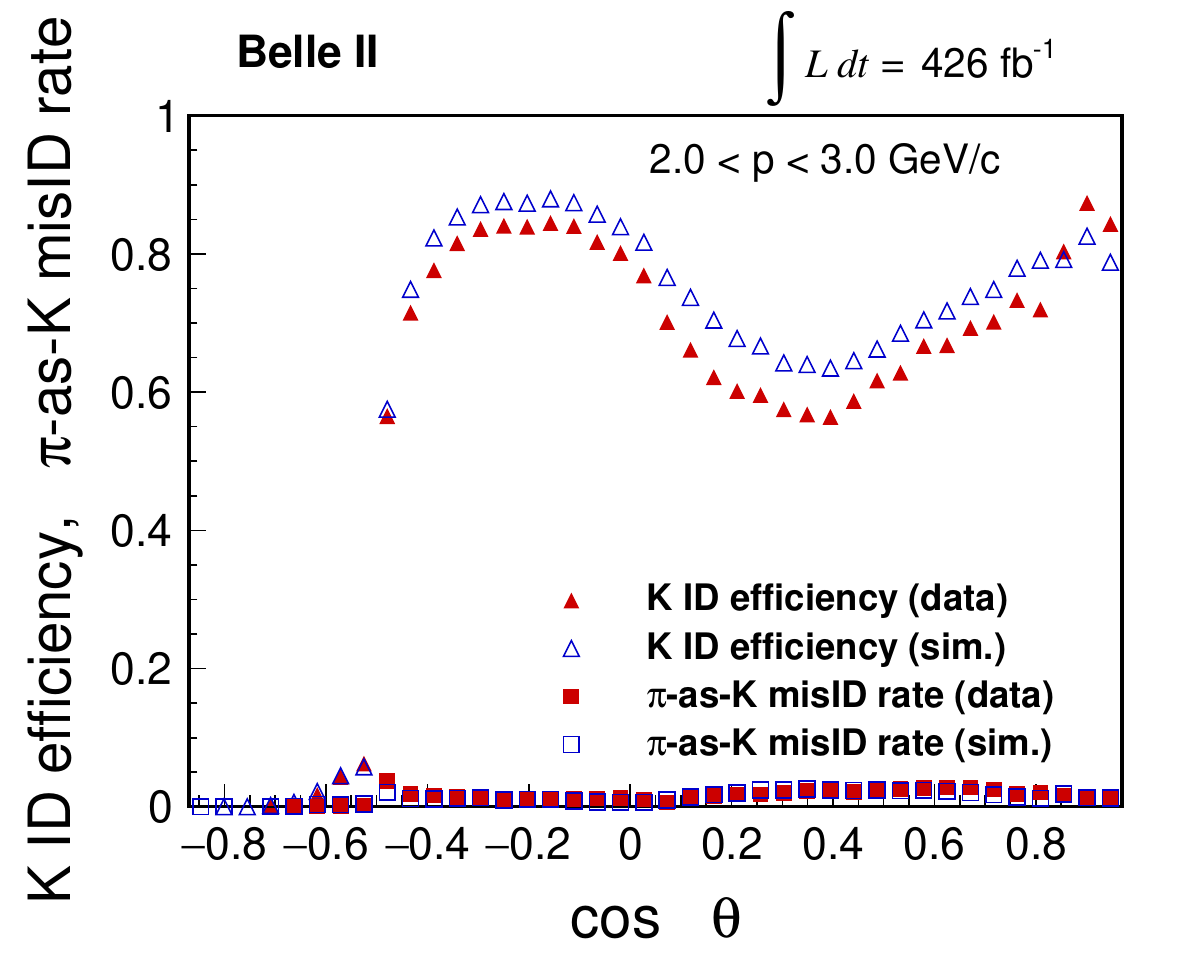}   \hfill%
        \includegraphics[width=\twocolumnplotwidth]{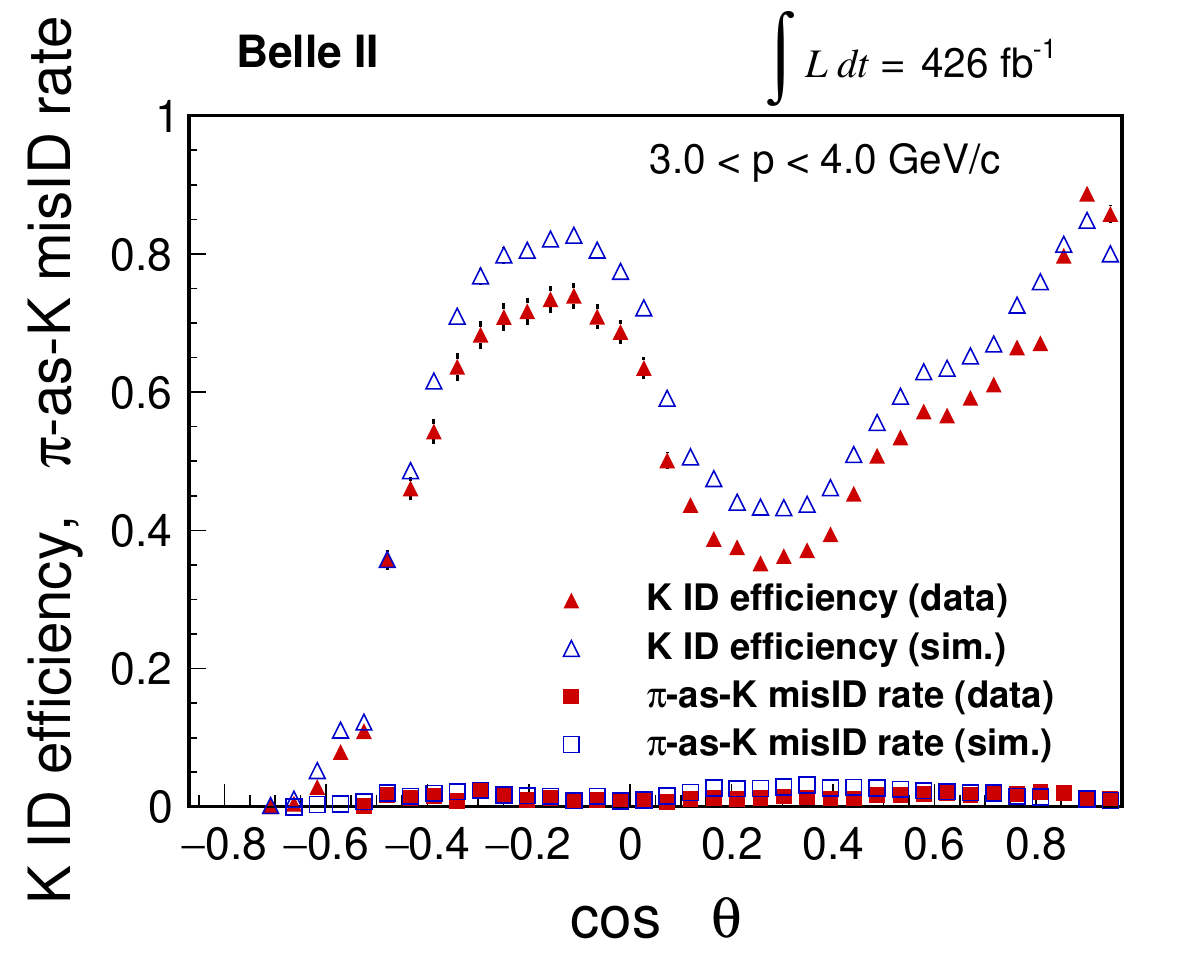} \\

        \caption{
            \PK-ID efficiency and \Ppi-as-\PK misID rate as functions of $\cos\theta$ for $P_{\PK/\Ppi} > 0.8$ using the neural-network probability in four momentum ranges.
        }
        
        \label{fig:eff-misID_costh_kpi_pbins}
    
    \end{center}
\end{figure*}

Fig.~\ref{fig:eff-misID_costh_kpi_pbins} shows the angular dependencies of the efficiency and misID rate (with $P_{\PK/\Ppi} > 0.8$ using the neural network) in four momentum ranges.
At low momenta, there is reasonable \PK-\Ppi separation even outside the TOP and ARICH acceptances, thanks to the CDC. 
At high momenta, there is a drop in separation power around $\cos\theta = 0.3$.
This stems from the TOP probability density functions for pions and kaons being very similar in that region of the phase space, from the imperfect correction of the chromatic error by the optics of the TOP modules, and from the fact that the number of detected photons reaches a minimum for that particular polar angle.
In the very forward region, particularly for $p\in [1.0, 2.0]$ GeV/c, there is a significant drop in performance, which originates from a drop in the ARICH and CDC performance in this momentum range.

\begin{figure*}[t!]
    \begin{center}

      \includegraphics[width=\textwidth]{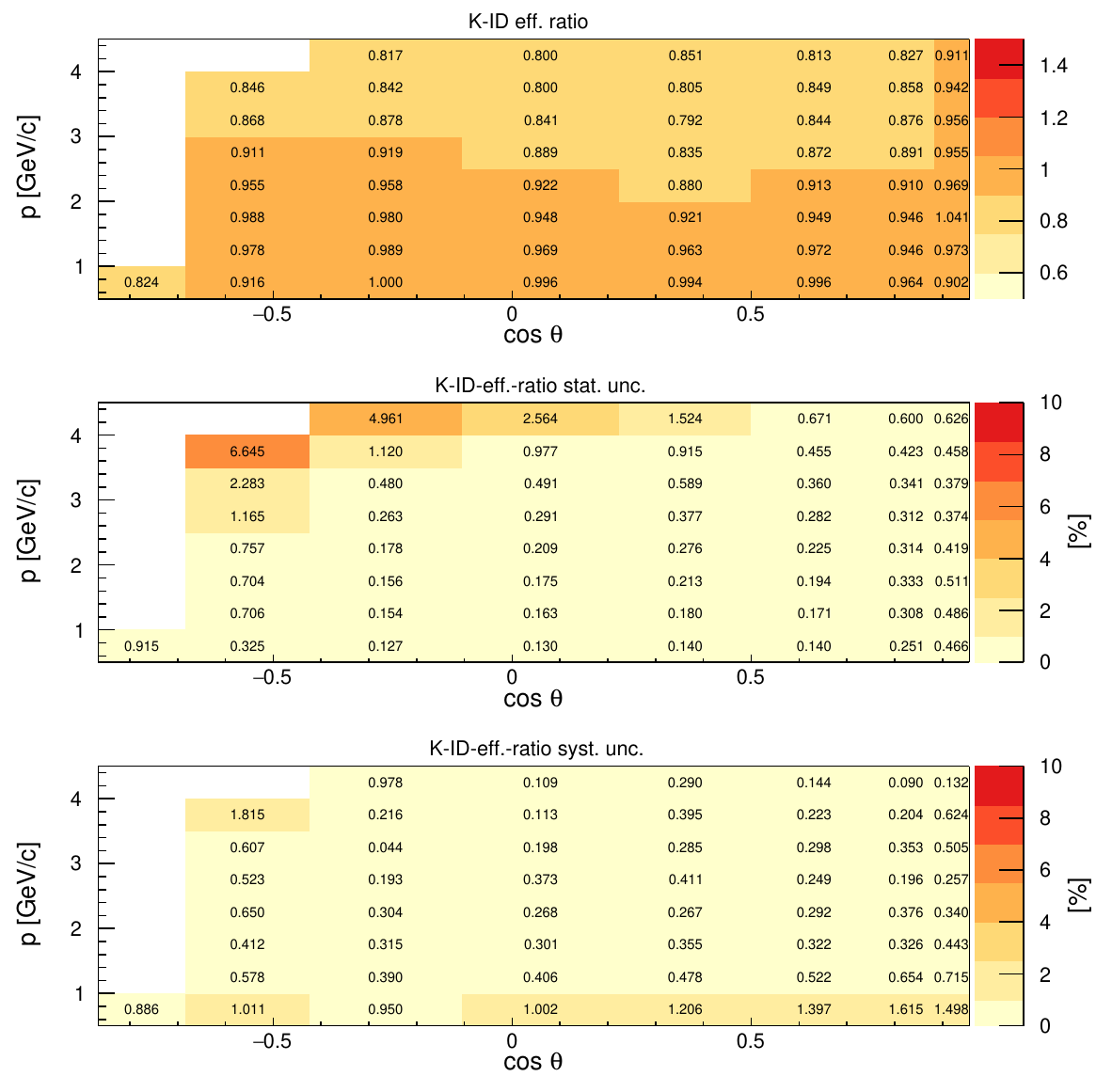}
      
        \caption{
            Ratio of the \PK-ID efficiency in data to that in simulation and its relative statistical and systematic uncertainties for $P_{\PK/\Ppi} > 0.8$ using the neural-network probability in subregions of $p$ and $\cos\theta$.
        }
        
        \label{fig:eff-ratio_kpi}

    \end{center}
\end{figure*}

Figure~\ref{fig:eff-ratio_kpi} shows the ratio of the \PK-ID efficiency in data to that in simulation and its relative statistical and systematic uncertainty for $P_{\PK/\Ppi} > 0.8$ using the neural-network probability in subregions of $p$ and $\cos\theta$. The binning in $\cos\theta$ reflects the physical boundaries between the TOP and ARICH and the ECL's barrel and end caps.

Each systematic uncertainty is the sum in quadrature of uncertainties from modeling of the background components in the fits that determine the sWeights in the control samples.
The main source of uncertainty is obtained by replacing the baseline linear fit used to model the background with a second-order polynomial.
We include a systematic uncertainty to account for potential bias from the sWeights procedure, which is the difference between the efficiency calculated for simulated events using sWeights and the efficiency calculated from knowing the true particle species used in the simulation.

In subregions where the statistical uncertainty is smallest---where most events are in data---the ratios are within 10\% of unity.
Except for very low momenta, the systematic uncertainties are comparable to the statistical ones.

\begin{figure*}[t!]
    \begin{center}

        \includegraphics[width=\textwidth]{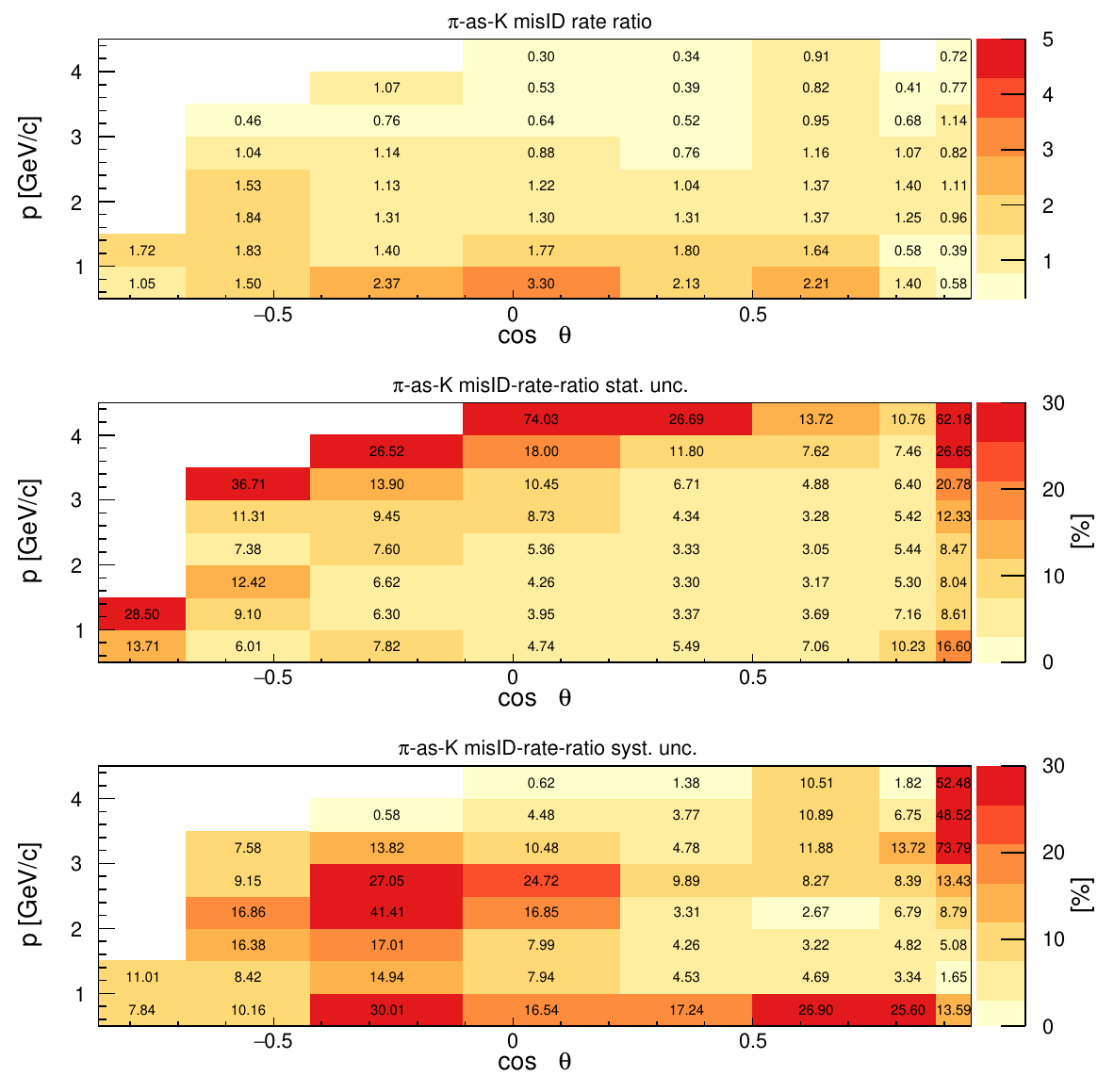}

        \caption{
            Ratio of the \Ppi-as-\PK misID rate in data to that in simulation and its relative statistical and systematic uncertainties for $P_{\PK/\Ppi} > 0.8$ using the neural-network probability in subregions of $p$ and $\cos\theta$.
        }

        \label{fig:misID-ratio_kpi}

    \end{center}
\end{figure*}

Figure~\ref{fig:misID-ratio_kpi} shows the ratio of the \Ppi-as-\PK misID rate in data to that in simulation for $P_{\PK/\Ppi} > 0.8$ using the neural-network probability.
This ratio deviates much more from one than the efficiency ratio.
Its systematic uncertainty, calculated in the same way as that for the efficiency ratio, is almost everywhere larger than the corresponding statistical uncertainty.
This stems from the relatively larger impact of background modeling on the misID rate, which is small, compared to the efficiency, which is large.

\subsection{\texorpdfstring{\Pp-\PK}{proton-kaon} separation}

\begin{figure*}[t!]
    \begin{center}

        \includegraphics[width=\twocolumnplotwidth]{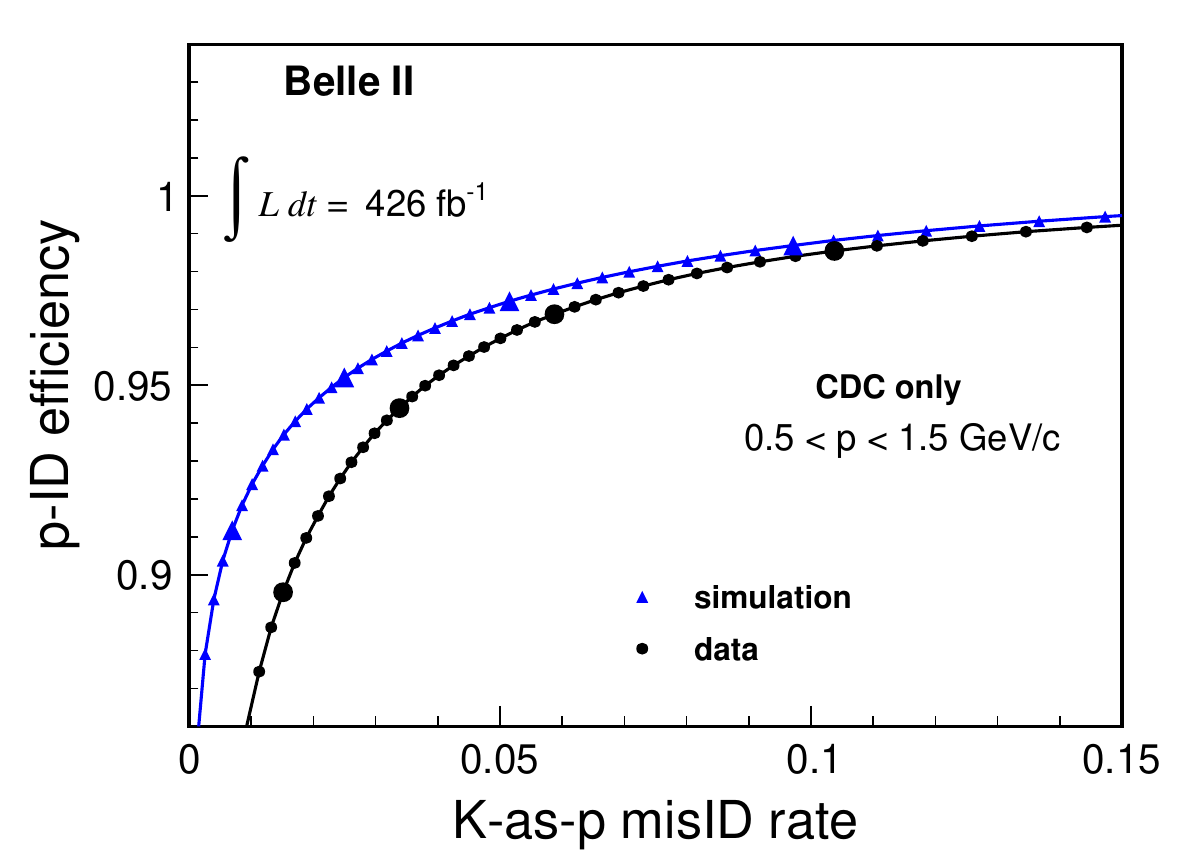} \hfill%
        \includegraphics[width=\twocolumnplotwidth]{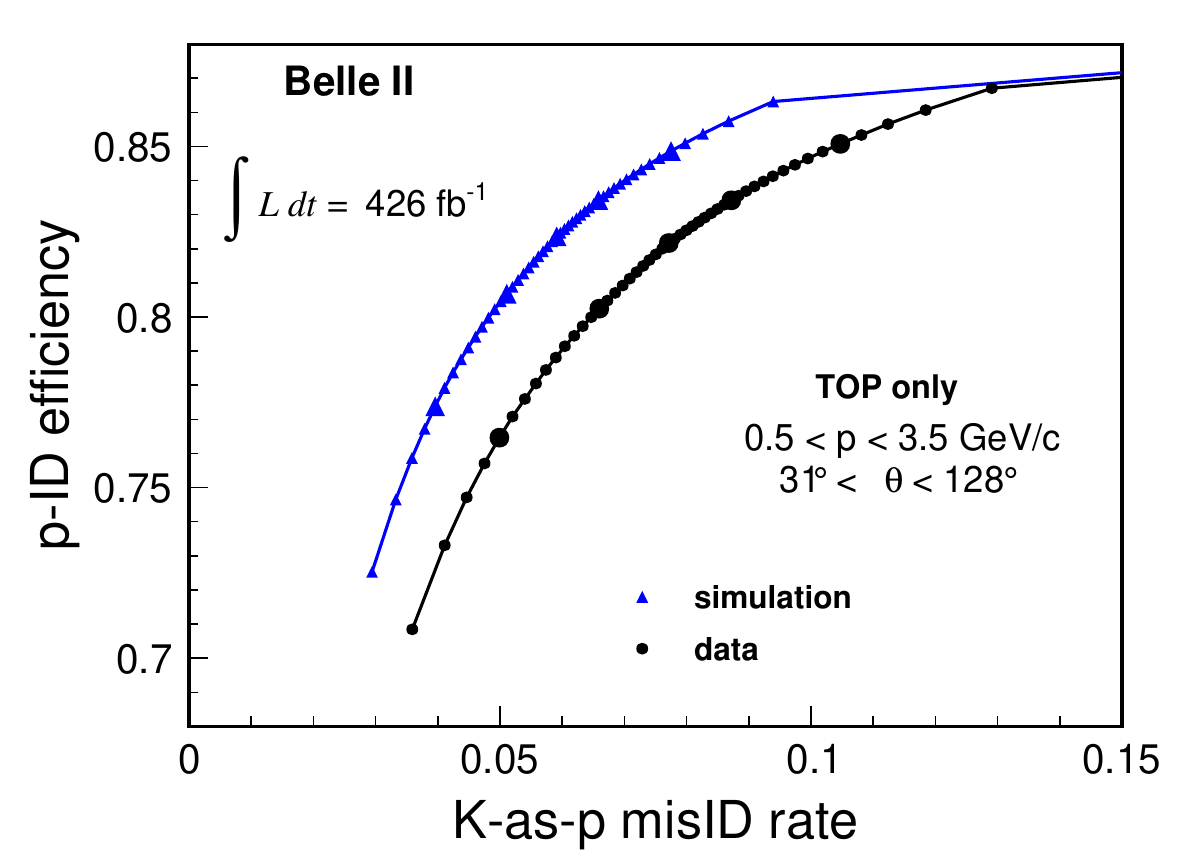}      \\
        \includegraphics[width=\twocolumnplotwidth]{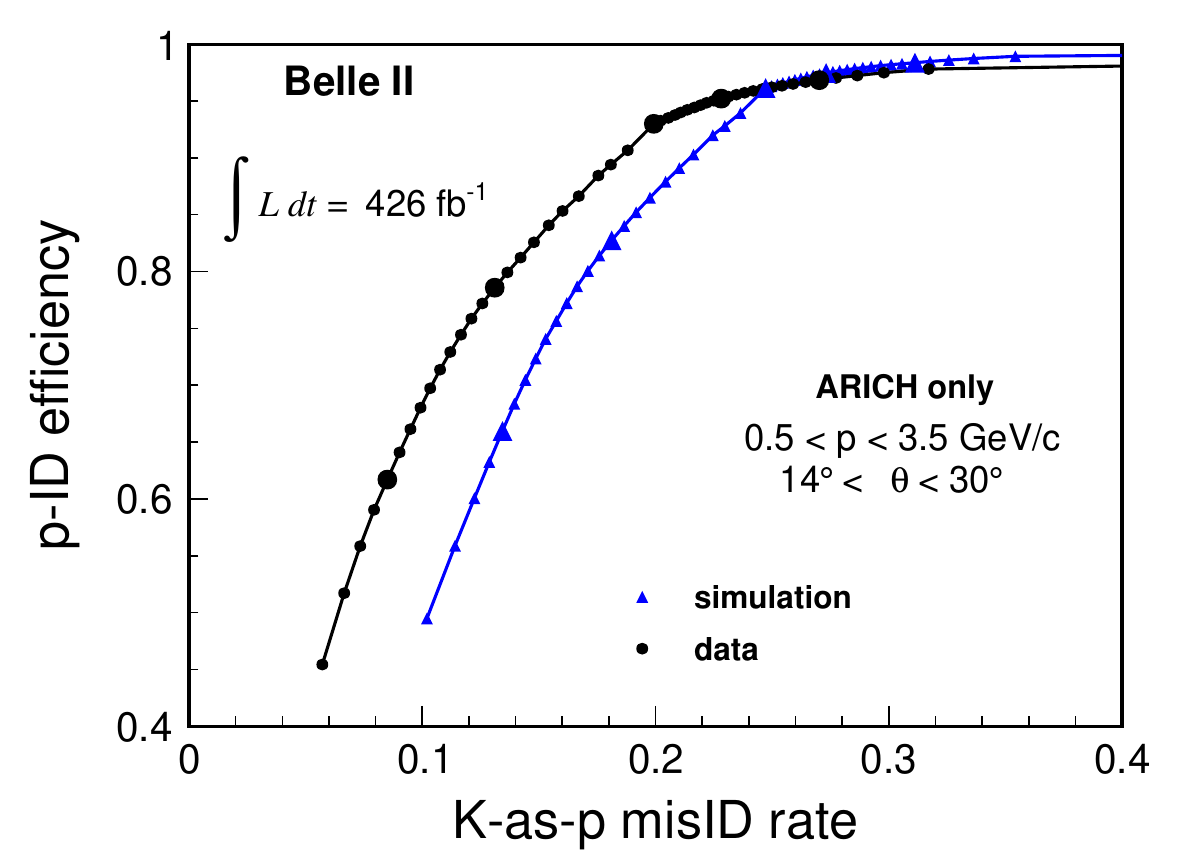}    \hfill%
        \includegraphics[width=\twocolumnplotwidth]{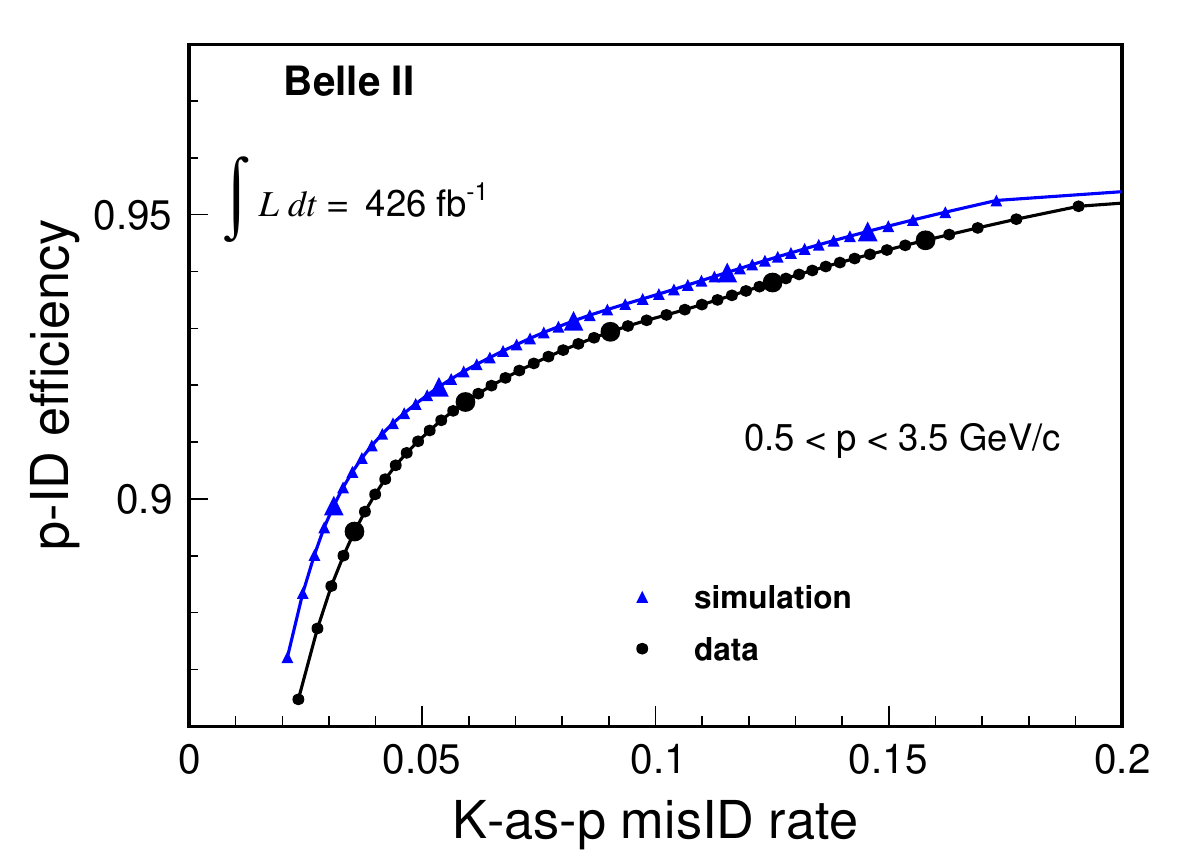}      \\

        \caption{
            \Pp-ID efficiency as functions of the \PK-as-\Pp misID rate for the local binary probabilities from the CDC~(for low momenta), TOP, and ARICH and using the simple binary probability with information from all subdetectors.
            The large markers represent the thresholds (from left to right) 0.9, 0.7, 0.5, 0.3, and 0.1; the lower thresholds are not always visible in the ranges shown.
        }

        \label{fig:eff-misID_local-likelihood_pk}

    \end{center}
\end{figure*}

We use the \PLambda control channel to calculate the \Pproton-ID efficiency and the \PDzero control channel to calculate the \PK-as-\Pproton misID rate.
Figure~\ref{fig:eff-misID_local-likelihood_pk} shows the \Pp-ID efficiency as functions of the \PK-as-\Pp misID rate for PID requirements using only $P^d_{\Pp/\PK}$ for $d = $ CDC, TOP, and ARICH, and for the combination of all subdetectors.

For the CDC, we show results for low momenta, as the CDC \Pp-\PK separation power is marginal at high momenta.
For the TOP and ARICH, we show results only for particles within their acceptance.
As for \PK-\Ppi separation, the CDC and TOP perform slightly worse in data than in simulation; the opposite is true for the ARICH.
Again, performance in real data and simulation agree better using all subdetectors.

\begin{figure*}[t!]
    \begin{center}
    
        \includegraphics[width=\twocolumnplotwidth]{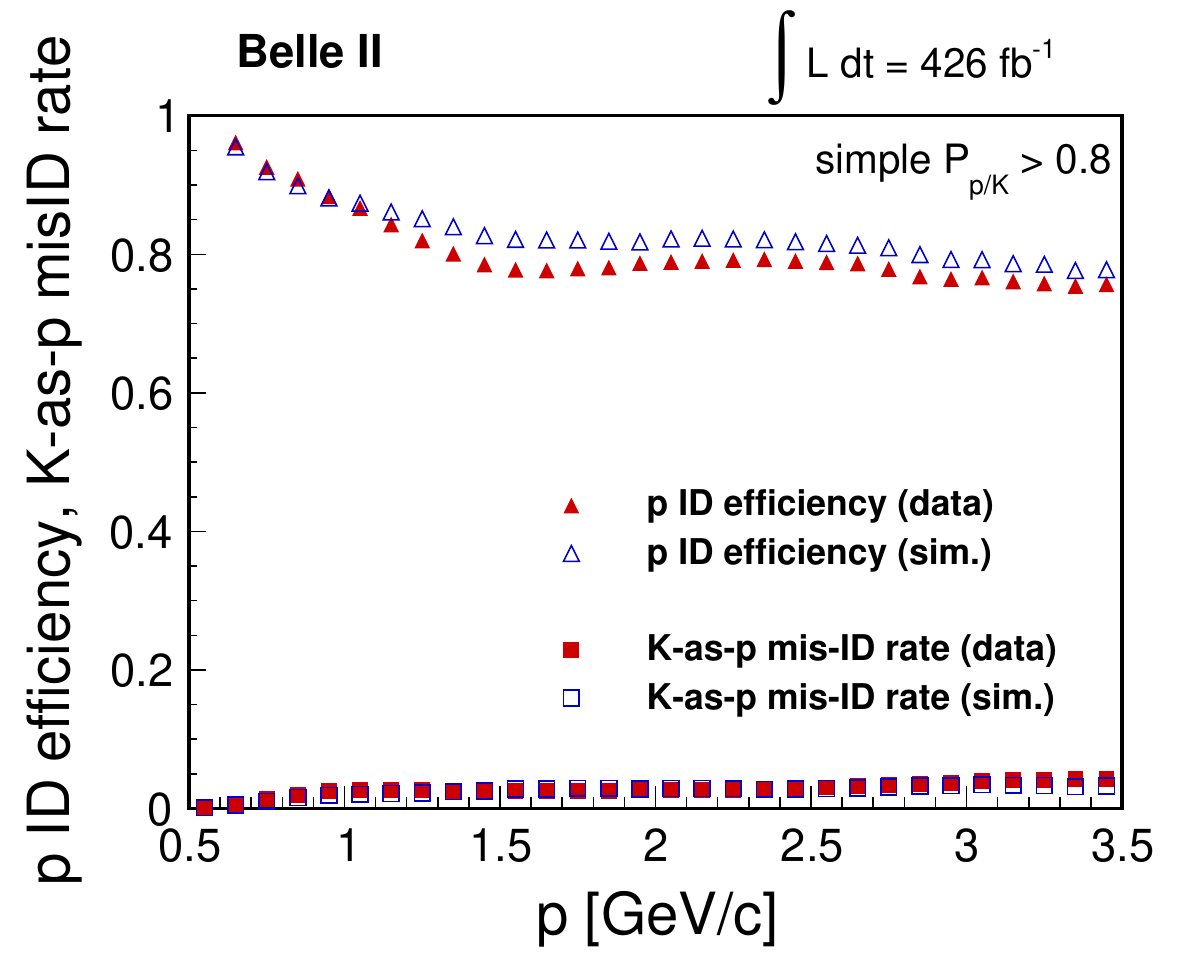}   \hfill%
        \includegraphics[width=\twocolumnplotwidth]{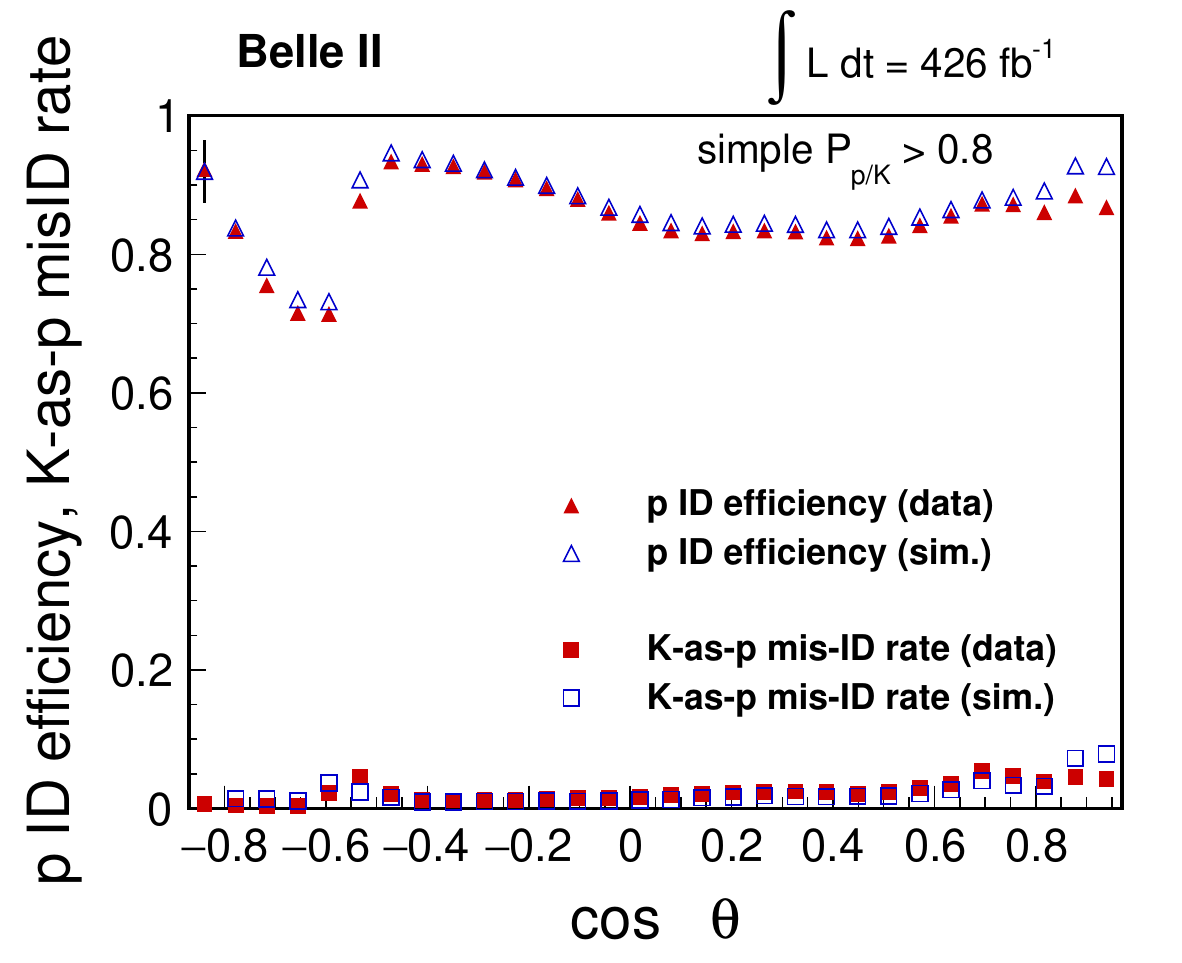} \\

        \caption{
            \Pp-ID efficiency and \PK-as-\Pp misID rate as functions of $p$ and $\cos\theta$ for $P_{\Pp/\PK} > 0.8$ using the simple binary probability.
        }

        \label{fig:eff-misID_pcosth_pk}

    \end{center}
\end{figure*}

\begin{figure*}[t!]
    \begin{center}
    
        \includegraphics[width=\twocolumnplotwidth]{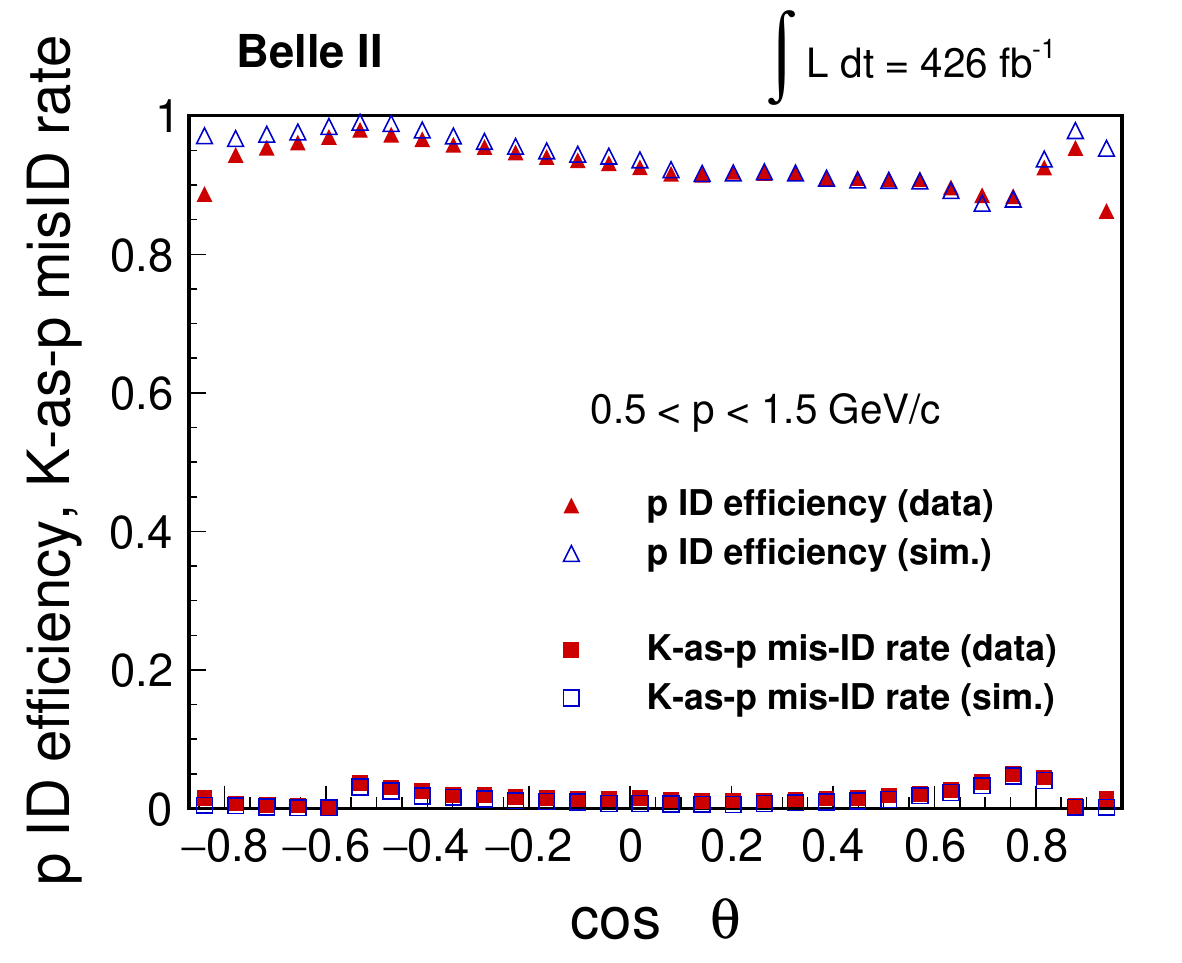}   \hfill%
        \includegraphics[width=\twocolumnplotwidth]{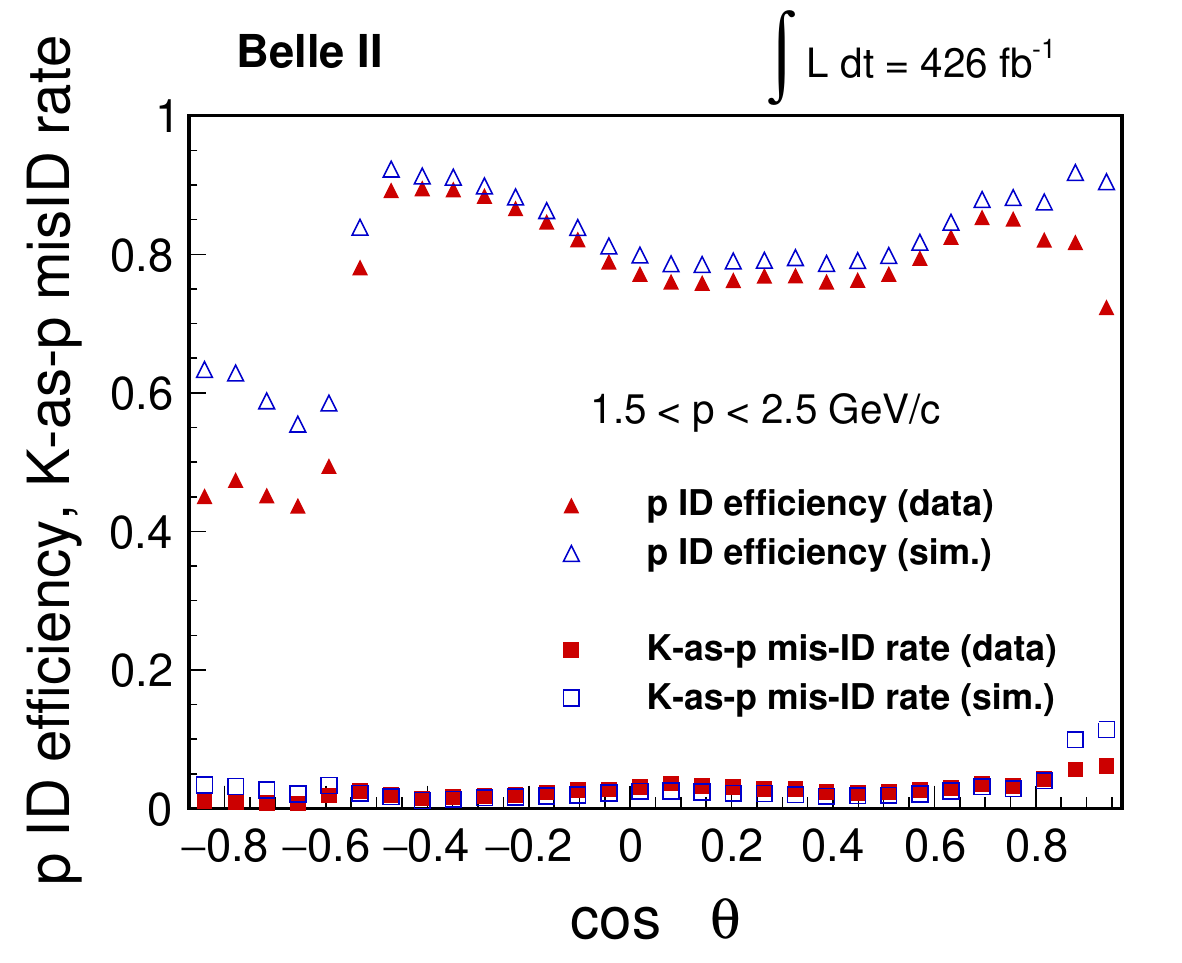} \\
        \includegraphics[width=\twocolumnplotwidth]{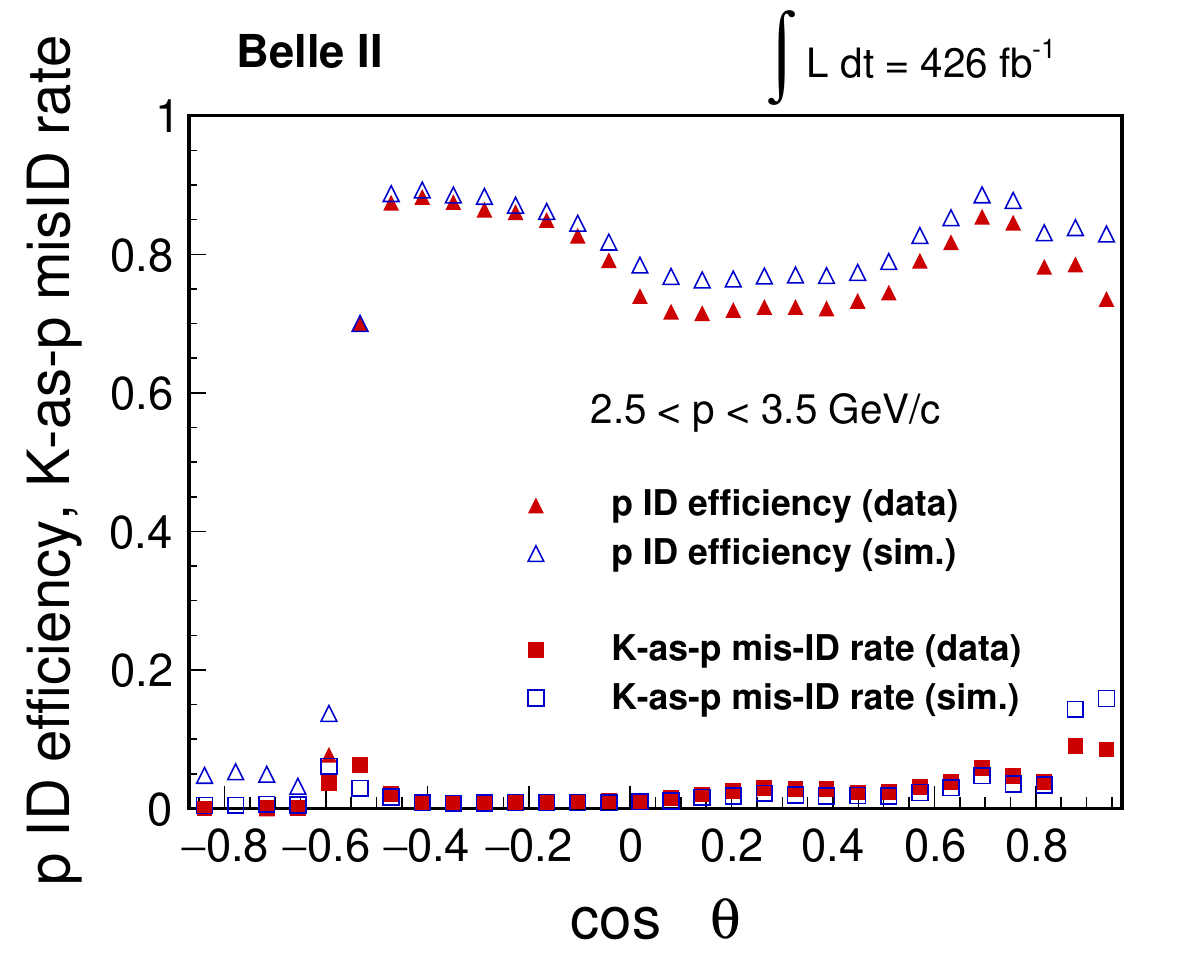}   \hfill%
        \quad

        \caption{
            \Pp-ID efficiency and \PK-as-\Pp misID rate as a function of $\cos\theta$ for $P_{\Pp/\PK} > 0.8$ using the simple binary probability in three momentum ranges.
        }

        \label{fig:eff-misID_costh_pk_pbins}

    \end{center}
\end{figure*}

\begin{figure*}[t!]
    \begin{center}

        \includegraphics[width=\textwidth]{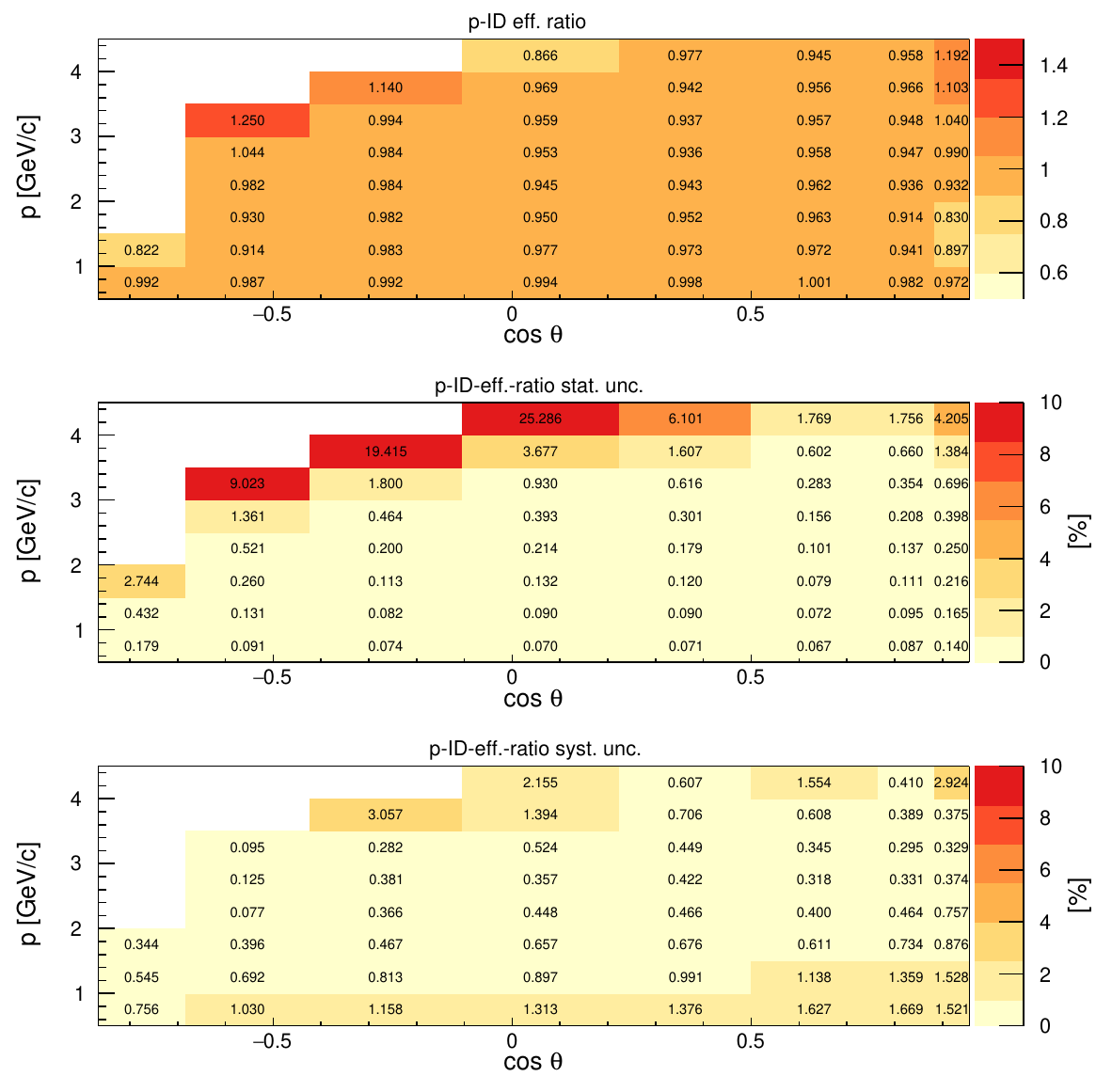}

        \caption{
            Ratio of the \Pp-ID efficiency in data to that in simulation and its relative statistical and systematic uncertainties for $P_{\Pp/\PK} > 0.8$ using the simple probabilities in subregions of $p$ and $\cos\theta$.
        }
        
        \label{fig:eff-ratio_pk}

    \end{center}
\end{figure*}

Figure~\ref{fig:eff-misID_pcosth_pk} shows the efficiency and the misID rate as functions of $p$ and $\cos\theta$ for $P_{\Pp/\PK} > 0.8$ for the simple binary probability in data and simulation, and Fig.~\ref{fig:eff-misID_costh_pk_pbins} shows both as functions of $\cos\theta$ for three momentum ranges.
At low momenta, the CDC is the most important subdetector for \Pp-\PK separation; the boundary of the TOP at $\cos\theta \sim -0.5$ is not even visible.
At high momenta, just as for \PK-\Ppi separation, the efficiency is relatively lower in the region of $\cos\theta = 0.3$.
Figure~\ref{fig:eff-ratio_pk} shows the ratio of the \Pp-ID efficiency in data to that in simulation for $P_{\Pp/\PK} > 0.8$ using the simple probability in subregions of $p$ and $\cos\theta$.
Data and simulation agree within 10\% in most subregions.

\subsection{\texorpdfstring{\Plepton-as-\Ppi}{lepton-as-pion} misID probability}

\begin{figure*}[t!]
    \begin{center}

        \includegraphics[width=\textwidth]{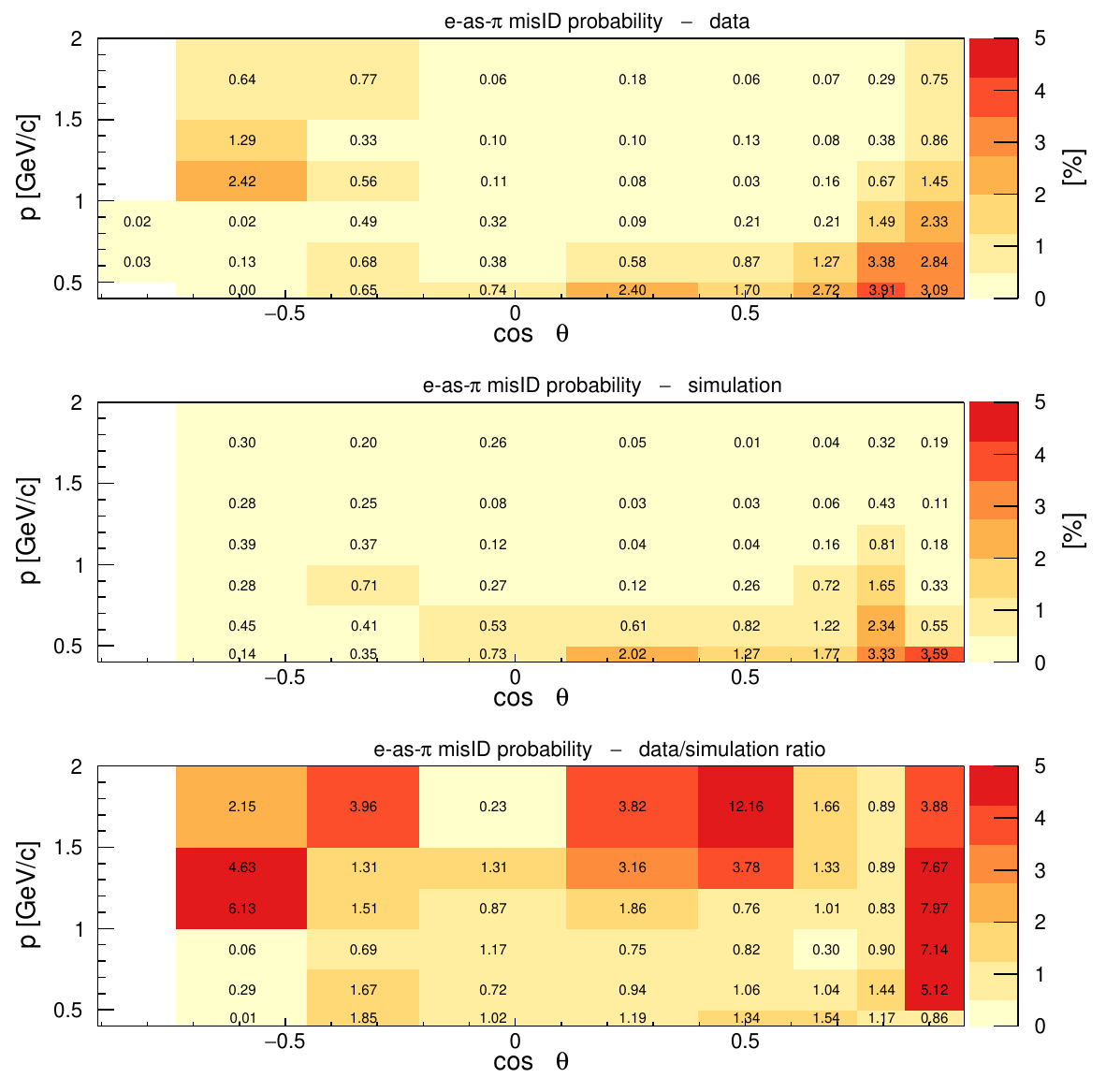}

        \caption{
            \Pe-as-\Ppi misID rate for $P_{\Ppi} > 0.8$ using the simple probability for subregions of $p$ and $\cos\theta$ in data~(top) and simulation~(middle) and their ratio~(bottom).
        }
      
        \label{fig:misID_pcosth_epi}
        
    \end{center}
\end{figure*}
\begin{figure*}[t!]
    \begin{center}
        
        \includegraphics[width=\textwidth]{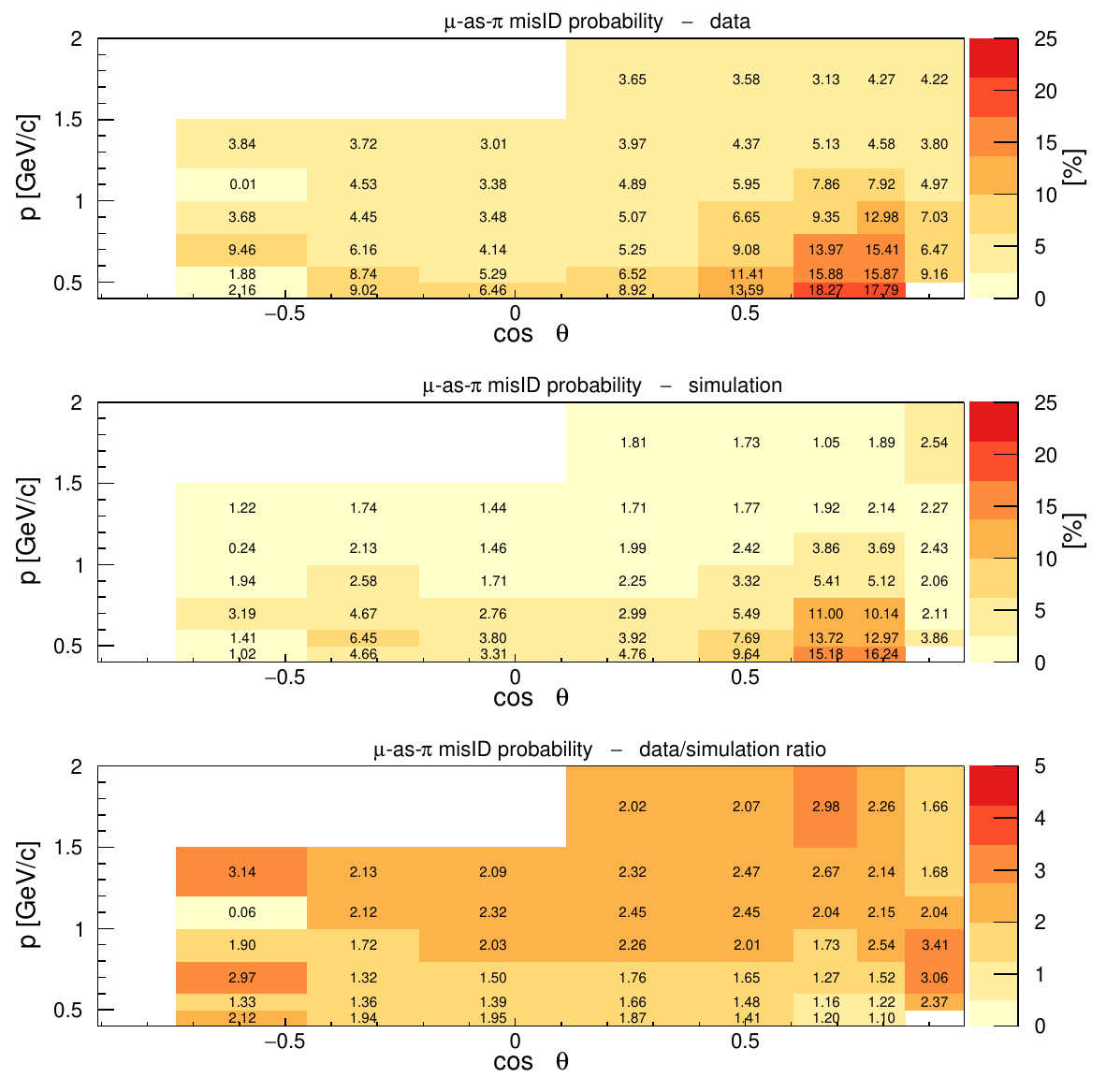}
       
        \caption{
            \Pmu-as-\Ppi misID rate for $P_{\Ppi} > 0.8$ using the simple probability for subregions of $p$ and $\cos\theta$ in data~(top) and simulation~(middle) and their ratio~(bottom).
        }
        
        \label{fig:misID_pcosth_mupi}
        
    \end{center}
\end{figure*}

We use four-lepton events to calculate the \Pe-as-\Ppi and \Pmu-as-\Ppi misID rates for $P_{\Ppi} > 0.8$ with the simple probability.
Figures~\ref{fig:misID_pcosth_epi} and~\ref{fig:misID_pcosth_mupi} show them in subregions of $p$ and $\cos\theta$ in data and simulation. In general, the results in data and simulation agree in their features. The data-simulation ratio, excluding some very forward and backward subregions, is below 3, which is acceptable for analyses.


%
%
%
\section{Future developments and conclusions}
\label{sec:future}

Work is in progress to improve particle identification, especially to mitigate the effects of the increased backgrounds that will accompany the higher instantaneous luminosity SuperKEKB will achieve in the future.

We can improve PID performance by optimizing calibration of the CDC to counteract performance degradation due to beam-injection-induced backgrounds and using machine learning in the reconstruction software for the TOP, ECL, and KLM. We are also extending the \PK-\Ppi neural network to identify all six particle species.

Though we did not quantitatively compare the PID performance of Belle II to that of the first-generation \PB-factory experiments, we can still make some general observations. From the performance plots available in~\cite{Bevan:2014iga}, we see that \BelleII's pion misID probability is 5\% lower than Belle's, which was around 7\%, at a \PK-ID effiency of 85\%, in the full acceptance of the detector and in the same control sample.
In the \PLambda control sample, \BelleII's \Pp-ID efficiency is high (and its \PK-as-\Pp misID rate low) over the whole momentum range, whereas Belle's drops steeply above \SI{1.5}{GeV/c}.
This most likely stems from the TOP and ARICH subdetectors, which contribute relevant information in the high-momentum region where the CDC cannot.
Compared to \mbox{\slshape B\kern-0.1em{\smaller A}\kern-0.1em B\kern-0.1em{\smaller A\kern-0.2em R}}, \BelleII's \PK-\Ppi separation is currently worse.
But our performance is still noteworthy given that \BelleII operates in harsher background conditions than those of Belle and \mbox{\slshape B\kern-0.1em{\smaller A}\kern-0.1em B\kern-0.1em{\smaller A\kern-0.2em R}}. 


This work, based on data collected using the Belle II detector, which was built and commissioned prior to March 2019,
was supported by
Higher Education and Science Committee of the Republic of Armenia Grant No.~23LCG-1C011;
Australian Research Council and Research Grants
No.~DP200101792, 
No.~DP210101900, 
No.~DP210102831, 
No.~DE220100462, 
No.~LE210100098, 
and
No.~LE230100085; 
Austrian Federal Ministry of Education, Science and Research,
Austrian Science Fund (FWF) Grants
DOI:~10.55776/P34529,
DOI:~10.55776/J4731,
DOI:~10.55776/J4625,
DOI:~10.55776/M3153,
and
DOI:~10.55776/PAT1836324,
and
Horizon 2020 ERC Starting Grant No.~947006 ``InterLeptons'';
Natural Sciences and Engineering Research Council of Canada, Compute Canada and CANARIE;
National Key R\&D Program of China under Contract No.~2024YFA1610503,
and
No.~2024YFA1610504
National Natural Science Foundation of China and Research Grants
No.~11575017,
No.~11761141009,
No.~11705209,
No.~11975076,
No.~12135005,
No.~12150004,
No.~12161141008,
No.~12475093,
and
No.~12175041,
and Shandong Provincial Natural Science Foundation Project~ZR2022JQ02;
the Czech Science Foundation Grant No. 22-18469S,  Regional funds of EU/MEYS: OPJAK
FORTE CZ.02.01.01/00/22\_008/0004632 
and
Charles University Grant Agency project No. 246122;
European Research Council, Seventh Framework PIEF-GA-2013-622527,
Horizon 2020 ERC-Advanced Grants No.~267104 and No.~884719,
Horizon 2020 ERC-Consolidator Grant No.~819127,
Horizon 2020 Marie Sklodowska-Curie Grant Agreement No.~700525 ``NIOBE''
and
No.~101026516,
and
Horizon 2020 Marie Sklodowska-Curie RISE project JENNIFER2 Grant Agreement No.~822070 (European grants);
L'Institut National de Physique Nucl\'{e}aire et de Physique des Particules (IN2P3) du CNRS
and
L'Agence Nationale de la Recherche (ANR) under Grant No.~ANR-21-CE31-0009 (France);
BMFTR, DFG, HGF, MPG, and AvH Foundation (Germany);
Department of Atomic Energy under Project Identification No.~RTI 4002,
Department of Science and Technology,
and
UPES SEED funding programs
No.~UPES/R\&D-SEED-INFRA/17052023/01 and
No.~UPES/R\&D-SOE/20062022/06 (India);
Israel Science Foundation Grant No.~2476/17,
U.S.-Israel Binational Science Foundation Grant No.~2016113, and
Israel Ministry of Science Grant No.~3-16543;
Istituto Nazionale di Fisica Nucleare and the Research Grants BELLE2,
and
the ICSC – Centro Nazionale di Ricerca in High Performance Computing, Big Data and Quantum Computing, funded by European Union – NextGenerationEU;
Japan Society for the Promotion of Science, Grant-in-Aid for Scientific Research Grants
No.~16H03968,
No.~16H03993,
No.~16H06492,
No.~16K05323,
No.~17H01133,
No.~17H05405,
No.~18K03621,
No.~18H03710,
No.~18H05226,
No.~19H00682, 
No.~20H05850,
No.~20H05858,
No.~22H00144,
No.~22K14056,
No.~22K21347,
No.~23H05433,
No.~26220706,
and
No.~26400255,
and
the Ministry of Education, Culture, Sports, Science, and Technology (MEXT) of Japan;  
National Research Foundation (NRF) of Korea Grants
No.~2021R1-F1A-1064008, 
No.~2022R1-A2C-1003993,
No.~2022R1-A2C-1092335,
No.~RS-2016-NR017151,
No.~RS-2018-NR031074,
No.~RS-2021-NR060129,
No.~RS-2023-00208693,
No.~RS-2024-00354342
and
No.~RS-2025-02219521,
Radiation Science Research Institute,
Foreign Large-Size Research Facility Application Supporting project,
the Global Science Experimental Data Hub Center, the Korea Institute of Science and
Technology Information (K25L2M2C3 ) 
and
KREONET/GLORIAD;
Universiti Malaya RU grant, Akademi Sains Malaysia, and Ministry of Education Malaysia;
Frontiers of Science Program Contracts
No.~FOINS-296,
No.~CB-221329,
No.~CB-236394,
No.~CB-254409,
and
No.~CB-180023, and SEP-CINVESTAV Research Grant No.~237 (Mexico);
the Polish Ministry of Science and Higher Education and the National Science Center;
the Ministry of Science and Higher Education of the Russian Federation
and
the HSE University Basic Research Program, Moscow;
University of Tabuk Research Grants
No.~S-0256-1438 and No.~S-0280-1439 (Saudi Arabia), and
Researchers Supporting Project number (RSPD2025R873), King Saud University, Riyadh,
Saudi Arabia;
Slovenian Research Agency and Research Grants
No.~J1-50010
and
No.~P1-0135;
Ikerbasque, Basque Foundation for Science,
State Agency for Research of the Spanish Ministry of Science and Innovation through Grant No. PID2022-136510NB-C33, Spain,
Agencia Estatal de Investigacion, Spain
Grant No.~RYC2020-029875-I
and
Generalitat Valenciana, Spain
Grant No.~CIDEGENT/2018/020;
The Knut and Alice Wallenberg Foundation (Sweden), Contracts No.~2021.0174 and No.~2021.0299;
National Science and Technology Council,
and
Ministry of Education (Taiwan);
Thailand Center of Excellence in Physics;
TUBITAK ULAKBIM (Turkey);
National Research Foundation of Ukraine, Project No.~2020.02/0257,
and
Ministry of Education and Science of Ukraine;
the U.S. National Science Foundation and Research Grants
No.~PHY-1913789 
and
No.~PHY-2111604, 
and the U.S. Department of Energy and Research Awards
No.~DE-AC06-76RLO1830, 
No.~DE-SC0007983, 
No.~DE-SC0009824, 
No.~DE-SC0009973, 
No.~DE-SC0010007, 
No.~DE-SC0010073, 
No.~DE-SC0010118, 
No.~DE-SC0010504, 
No.~DE-SC0011784, 
No.~DE-SC0012704, 
No.~DE-SC0019230, 
No.~DE-SC0021274, 
No.~DE-SC0021616, 
No.~DE-SC0022350, 
No.~DE-SC0023470; 
and
the Vietnam Academy of Science and Technology (VAST) under Grants
No.~NVCC.05.12/22-23
and
No.~DL0000.02/24-25.

These acknowledgements are not to be interpreted as an endorsement of any statement made
by any of our institutes, funding agencies, governments, or their representatives.

We thank the SuperKEKB team for delivering high-luminosity collisions;
the KEK cryogenics group for the efficient operation of the detector solenoid magnet and IBBelle on site;
the KEK Computer Research Center for on-site computing support; the NII for SINET6 network support;
and the raw-data centers hosted by BNL, DESY, GridKa, IN2P3, INFN, 
and the University of Victoria.

\bibliography{belle2}

\end{document}